\documentclass[aps,prf]{revtex4}

\usepackage{graphicx}

\usepackage{amsmath,amsfonts,bm, color,amssymb,float}
\usepackage{graphicx,epsfig,overpic,hyperref,ulem,caption,subcaption}

\newcommand{\out}{{\mathrm{out}}}
\newcommand{\ins}{{\mathrm{ins}}}

\newcommand{\sgn}{{\text{ sign}}}

\newcommand{\surf}{{\mathrm{s}}}
\newcommand{\sd}{{\mathrm{SD}}}

\newcommand{\Ca}{\mbox{\it Ca}}

\newcommand{\bomega}{{\bm {\omega}}}

\newcommand{\bv}{{\bm{u}}}
\newcommand{\bV}{{\bm{U}}}

\newcommand{\bu}{{\bm{u}}}

\newcommand{\s}{{\sigma}}

\newcommand{\Rdomain}{R_0}

\newcommand{\xhat}{{\bf \hat x}}

\newcommand{\zhat}{{\bf \hat z}}
\newcommand{\rhat}{{\bf \hat r}}
\newcommand{\phat}{{\bf \hat p}}

\newcommand{\phihat}{{\bm{ \hat \phi}}}
\newcommand{\nhat}{{\bf \hat n}}
\newcommand{\bx}{{\bf x}}
\newcommand{\bxz}{{\bf x_0}}

\newcommand{\domain}{d}
\newcommand{\particle}{p}

\newcommand{\bnab}{{\bm{\nabla}}}

\newcommand{\undisplaced}{{\mathrm{act}}}
\newcommand{\displaced}{{\mathrm{act}}}
\newcommand{\visc}{\eta}

\newcommand{\charforce}{F}

\newcommand{\bT}{{\bf{T}}}

\newcommand{\summs}{\sum_{m,\s}}
\newcommand{\summsp}{\sum_{m',\s'}}

\newcommand{\steadyshape}{\mathrm{ss}}
\newcommand{\stresslet}{\mathrm{str}}
\newcommand{\dipole}{\mathrm{dip}}

\newcommand{\refeq}[1]{Eq. (\ref{#1})}
\newcommand{\reffig}[1]{Fig. (\ref{#1})}
\newcommand{\refsec}[1]{Section (\ref{#1})}
\newcommand{\refapp}[1]{Appx. (\ref{#1})}

\begin{document}

\title{Activity-driven motion and deformation of a circular membrane domain}

\author{Gonzalo Ferrandez Quinto, Sho Kawakami and  Petia M. Vlahovska}
\affiliation{Engineering Sciences and Applied Mathematics, Northwestern University, Evanston, IL 60208, USA}

\begin{abstract}

Lipid bilayer membranes are nanometer-thin fluid interfaces characterized by both in-plane fluidity and lateral compositional heterogeneity. Cholesterol-rich liquid domains ("lipid rafts") are thought to regulate the spatial organization of membrane proteins and numerous cellular processes. While the passive diffusion and equilibrium shape fluctuations of membrane domains are well understood, much less is known about how the activity of molecular machines, such as protein pumps and enzymes, influences their mobility and deformation. Here, we analyze the dynamics of a membrane domain driven by an embedded molecular machine within the framework of two-dimensional Stokes flow. The molecular machine is modeled as a point singularity of the two-dimensional Stokes equations, including both rotational and irrotational singularities corresponding to force and source multipoles. We show that, despite being force-free, both force and source dipoles generate active flows that propel the membrane domain.  For a force dipole undergoing rotational diffusion, the domain motion is diffusive at long times but exhibits a transient superdiffusive regime at intermediate times.

\end{abstract}

\maketitle


\section{Introduction}

Biological membranes that enclose cells and cellular organelles are nanometer-thin lipid bilayers containing embedded proteins that regulate cellular communication and mediate the transport of ions and molecules across the membrane. These functions rely on the membrane's fluidity, which permits the lateral motion of proteins and other membrane inclusions \citep{lenaz1987lipid, Marguet:2006,Block2018Brownian,mogre2020getting}. Protein mobility was first analyzed in the seminal works of \citet{saffman1975brownian,saffman1976brownian} and has remained the subject of sustained interest ever since \citep{SHI2026}. 

Saffman and Delbrück modeled a membrane protein as a cylindrical inclusion embedded in a thin viscous sheet. Because translational motion in an unbounded two-dimensional Stokes flow is subject to the Stokes paradox, they employed a matched asymptotic analysis in which the near field flow is governed by the two-dimensional Stokes equations within the membrane, while  dissipation in the surrounding fluid becomes important beyond a distance (the Saffman--Delbrück length), $\ell_{\mathrm{SD}}=\eta_m/(\eta_{+}+\eta_{-})$, where $\eta_m$ is the membrane 2D viscosity and $\eta_\pm$ are the viscosities of the surrounding fluids. In the limit $a/\ell_{\mathrm{SD}}\ll1$, they showed that the mobility depends logarithmically on the protein radius $a$. Subsequently, \citet{hughes1981translational} extended the analysis numerically to arbitrary values of $a/\ell_{\mathrm{SD}}$.
 An alternative regularization of the Stokes paradox in this problem, also considered by \citep{saffman1975brownian,saffman1976brownian}, is based upon the finite extent of the membrane, imposing vanishing  velocity at a finite distance from the inclusion. Their analysis considered a circular inclusion positioned concentrically within a much larger circular boundary. More recently, \cite{yariv2025mobility} generalized this bounded configuration to an eccentric inclusion and examined how its position relative to the outer boundary affects the diffusivity. A relevant experimental realization of such a finite membrane geometry was provided by \cite{schutte2017size}, who investigated the size-dependent diffusion of inclusions in finite membrane patches.

Membrane-residing proteins such as ion pumps, including bacteriorhodopsin \citep{Manneville:1999}, and rotary molecular motors such as ATP synthase \citep{Almendro-Vedia11291,oppenheimer2019rotating}, perform mechanical work by converting chemical energy into motion without external forcing. As such, these molecular machines constitute biological active particles \citep{mavroidis2004molecular}. Membrane fluctuations (shape undulations) driven by active forces have attracted considerable attention as reporters of non-equilibrium metabolic activity in living cells \citep{Manneville:2001,Turlier:2019} and as mechanisms underlying cellular shape transformations \citep{sciortino2025active}.
More recently, the dynamics of active particles embedded in membranes have emerged as a topic of growing interest \citep{oppenheimer2019rotating,Hari:2020,shoham2023hamiltonian,manikantan2024stability,Mahapatra2025ActiveRods}, owing to their potential role in the self-organization of membrane inclusions and the emergence of membrane compartmentalization. A variety of  collective behaviors have been predicted and observed, including fractal-like clustering \citep{bashan2024hydrodynamically}, the formation of ordered hexagonal assemblies \citep{oppenheimer2019rotating}, and enhanced diffusion of passive tracers induced by protein activity 
\citep{hosaka2017lateral}.

Membrane proteins often localize within lipid rafts \citep{Simons1997, lingwood2010lipid}, which are membrane domains enriched in cholesterol and sphingolipids that are more ordered and tightly packed than the surrounding membrane, while remaining fluid \citep{veatch:2002,membranes11050323}; biomolecules themselves can  trigger raft formation \citep{gao2025polymer}. 
 Rafts can be viewed as droplets within a two-dimensional fluid membrane and are typically assumed to diffuse passively \citep{Cicuta:2007}.  
An important open question, however, is whether activity-generated flows produced by protein machines confined to a raft can influence the raft  motion, interactions, and stability.  Active particles enclosed by a soft boundary in two dimensions have been studied using a spring-bead model for the boundary and the active Brownian particle model for the particle motion in \citep{paoluzzi2016shape,wang2019shape}  and within a phase-field model \citep{diaz2024active}, and shown to give rise to diffusive motion of the droplet. Notably, these studies 
neglect hydrodynamic flows. However, recent work examining the dynamics of a droplet enclosing active particles \citep{kokot2022spontaneous,kawakami2025migration, kawakami2025microswimmer} accounting for the fluid flows stirred by the particle activity and the hydrodynamic interaction with the confining boundary demonstrated persistent droplet motion. Could activity also drive superdiffusive motion of domains in membranes?

In this work, we analyze the flow due to  an active particle modeled as a point Stokes-flow singularity in a two-dimensional fluid. Protein machines operate in a highly viscous environment where inertia is negligible \citep{astumian2007design}. Under such conditions, forces are locally balanced, so any force exerted by the machine on the fluid is accompanied by an equal and opposite reaction on its surroundings. Consequently, molecular machines often modeled as force dipoles \citep{Manneville:2001,mikhailov2015hydrodynamic}.

We develop an analytical solution for the two-dimensional flow driven by an arbitrary 
 Stokes singularity inside a domain modeled as a droplet. 
We find that even a force-free singularity such as the force dipole can drive domain translation. 
Domain deformation is analyzed in the limit of small deviations from the equilibrium circular shape,  where the line (edge) tension is stronger than the viscous stresses due to the active flow. 
Our findings complement the recent studies in three dimensions \citep{kawakami2025migration,daddi2018creeping,aponte2022confined,sun2025dynamics} and Hele-Shaw configurations \citep{kawakami2025microswimmer}.

\section{Problem formulation}

\label{sec:Formulation}
{The typical size of a transmembrane protein is approximately $5\,\mathrm{nm}$, while lipid-raft sizes span a broad range \citep{sezgin2017mystery,levental2020lipid,Chuangsciadv2025}. In vivo, membrane domains are generally estimated to be in the range $10-200\,\mathrm{nm}$ \citep{pike2006rafts}, whereas in vitro systems can exhibit domains extending to micrometer scales \citep{veatch:2002,Veatch-Keller:2005,mangiarotti2026phase}. These length scales are typically smaller than the Saffman--Delbrück length $\ell_\sd$. For a membrane viscosity of $1\text{--}10\times10^{-9}\,\mathrm{N.s/m}$ \citep{Faizi2022} and water as the surrounding fluid, $\ell_\sd\sim 0.5\text{--}5\,\mu\mathrm{m}$. Thus, for domain radii $R_0\ll\ell_\sd$,  dissipation is mainly in the membrane and the flow can be approximated as two-dimensional. 

Henceforth, all variables are nondimensionalized by characteristic length $\Rdomain$ and time $\eta_m/\tau_c$, respectively. For a force dipole, let $F$ denote the magnitude of each force and ${\cal P}=Fd$ the dipole strength, where $d$ is their separation and is of order the protein size. The corresponding characteristic active stress is $\tau_c={\cal P}/R_0^2$. 
Characteristic stresses for the other singularities are defined analogously from their respective strengths. 

}

\begin{figure}
\centering
\includegraphics[width=0.5\textwidth]{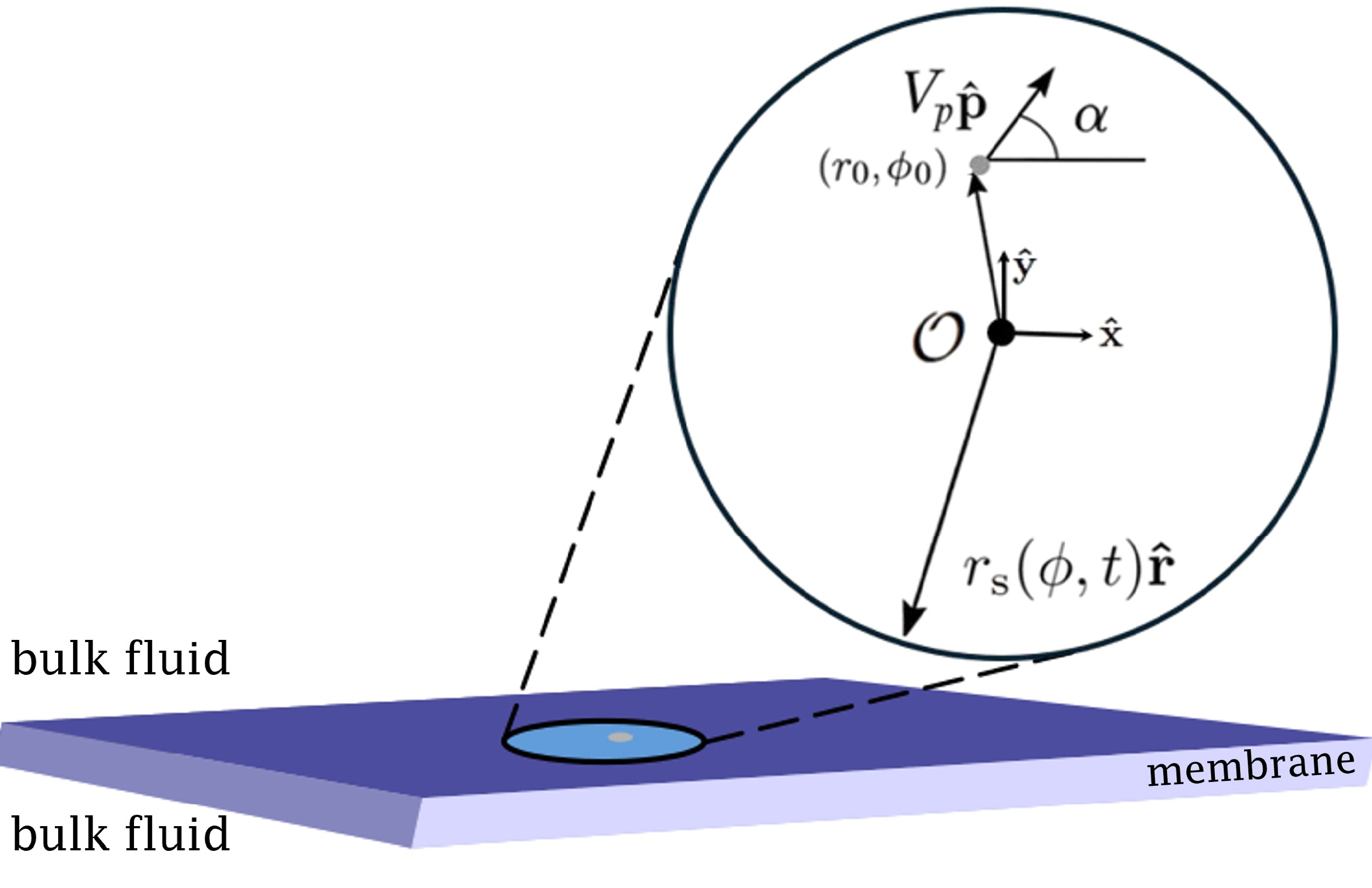}
\caption{{\textbf{Problem setup:}} A {quasi-}  circular domain of viscosity $\lambda \eta_m $ in a planar membrane of viscosity $\eta_m$. Inside the domain, an active particle located at $\textbf x_0 =(r_0,\phi_0)$ moves with active velocity $V_\particle$ in direction $\phat$. The vector $\phat$ forms an angle $\alpha$ with the x-axis. The domain boundary is free to evolve {in response to the flow generated by the active particle}  and the domain shape  is parametrized by $r_\surf(\phi,t)$.}
\label{fig Diagram}
\end{figure}

The fluid velocity and pressure inside the domain, $\bv^{\ins}$ and $p^{\ins}$, and outside the domain, $\bv^{\out}$ and $p^{\out}$, are described by the Stokes equation in non-dimensional form
\begin{align}
-\bnab  p^{\out}+\nabla^2 \bv^{\out} = 0\,, &\quad{\bnab \cdot \bv^{\out} = 0},\label{eq: StokesEqoutside}
\\
-\bnab  p^{\ins}+\lambda\nabla^2 \bv^{\ins} = \bm{f(\bx-\bxz)}\,, & \quad{\bnab \cdot \bv^{\ins} = g(\bx-\bxz)} 
\label{eq: StokesEqinside}.
\end{align}
{$\lambda$ is the ratio of domain to surrounding membrane viscosities, typically  $\lambda\sim 5$ \citep{schutte2017size}.} 
The point distributions $\bm{f}(\bx-\bxz)$ and $g(\bx-\bxz)$, {which respectively correspond to force and potential flow multipoles}, model the active particle located at $\bxz$ in a coordinate system centered at the domain origin; see \reffig{fig Diagram} for a sketch of the problem setup.
{
Although molecular machines are naturally represented by force multipoles, we also consider source multipoles, whose irrotational velocity fields provide additional solutions of the Stokes equations. Note that, except the Stokeslet, the force and source multipoles considered here generate flows that decay at infinity and therefore do not encounter the Stokes paradox associated with two-dimensional translation.}

The position of the domain interface is specified by $\bx_\surf = r_\surf(\phi,t)\rhat$, with $\phi$ being the polar angle. 
It evolves according to 
\begin{equation}
       \frac{d \bx_\surf}{dt} = \bv_\surf\,,
       \label{Eq: surface evolution}
\end{equation}
where $\bv_\surf$ is the interface velocity. The velocity is continuous across the domain interface
\begin{equation}
\bv^{\ins} = \bv^{\out} = \bv_\surf, \,\, \text{at }r=r_\surf  \label{BC: continuity}.
\end{equation}
The bulk hydrodynamic stress is given by,
\begin{equation}
\bm{T}^{(i)} = -p^{(i)}\bm{I}+\eta^{(i)}\left[\bnab \bv^{(i)}+(\bnab\bv^{(i)})^T\right], \label{Eq: bulkstress}
\end{equation}
with $(i) \in \{\ins, \out\}$ and $\eta^\out=1,\eta^\ins=\lambda$. The jump in normal stress across the domain interface is balanced by the line tension, expressed as
\begin{equation}
\nhat \cdot (\bm{T}^{\out}-\bm{T}^{\ins}) = \Ca^{-1} \left(\bnab \cdot\nhat \right)\nhat,\,\,\,r=r_\surf \label{BC: normal stress balance},
\end{equation}
where $\Ca = \frac{\tau_c R_0}{\gamma}$ is the capillary number, quantifying the relative strength of the singularity-driven viscous flow stresses to the line tension. 

We consider the general case of a self-propelled active particle, motivated by recent advances in engineered nanomotors \citep{ju2025technology}.
The active particle's position $\bxz$ evolves due to self-propulsion with intrinsic velocity ${v_\particle}$ along the direction $ \phat$, and it is also advected and reoriented by the surrounding fluid flow. The particle's trajectory is given by
\begin{equation}
\begin{split}
&\frac{d \bm{\bxz}}{dt} = {V}_\particle \phat+\bv_\particle , \,\,\quad \frac{d\phat}{dt} = \bomega_\particle \times \phat\,,
\end{split}\label{Eq: Particle Evo Eq}
\end{equation}
where the normalized particle velocity is ${V}_\particle = v_\particle \visc  /\charforce$, $\bv_\particle$ is the local flow velocity, and $\bomega_\particle=\frac{1}{2}\nabla\times \bu_\particle$ denotes the rotation rate. {We neglect alignment by the rate of strain}.

\section{Asymptotic solution for small shape deformation}\label{sec: Solution Method}
{An active particle within the domain generates a flow,
which
deforms the domain while the line tension acts to keep the domain shape circular. The relative strength of the shape-deforming active forcing and the shape-restoring line tension is quantified by the capillary number, $\Ca$. Molecular machines typically exert forces on the picoNewton scale \citep{lipowsky:2020,hess2024}, and typical line tensions are in the order of $1\,\mathrm{pN}$ \citep{Baumgart:2003,Tian_Baumgart:2007}, hence $\Ca\sim 10^{-3}$ for micrometer-sized domains.}

An analytical solution of the problem is possible in terms of a regular perturbation
in the case of the weak activity, $\Ca \ll 1$. Under this condition,  the domain shape remains quasi-circular
\begin{equation}
r_\surf(\phi,t) = 1+\epsilon f(\phi,t)\,,\label{Eq: radial basis}
\end{equation}
where $\epsilon\equiv\Ca$ is the small parameter for the perturbation.
The interface position is expanded in a Fourier series 
\begin{equation}
    f(\phi,t) = \sum_{m=-\infty}^\infty f_{m}(t) e^{i m \phi}.
\end{equation}
Note that the mode $f_0$ corresponds to changes in the domain area. Since the membrane and domain fluids are assumed  incompressible and  there is no net source or sink within the domain, $f_0=0$. Similarly, the modes $f_{\pm1}$ are also set to zero since the reference frame is centered and co-moving with the domain.

\subsection{Solution for the fluid velocity}
The fluid flow inside and outside the domain is represented as a superposition of velocity modes $\bv_{m,\s}^\pm(\bx)$, which are the basis solutions of the 2D Stokes equation \citep{finken2008two}. The superscript $\pm$ denotes solutions that are regular either at infinity $(-)$ or at the origin $(+)$, and the index $\s$ distinguishes between irrotational $(\s=0)$ and rotational $(\s=1)$ modes. The explicit expressions for the basis velocity fields and the corresponding pressure fields are provided in \refapp{appendix: 2D singularity solutions}.

The  solution for the flow is found in three steps.

\begin{enumerate}
\item 
In free space, the active flow solution given by \refeq{eq: StokesEqinside} in a coordinate system centered at the singularity is expressed in decaying solutions of the Stokes equation
\begin{equation}
\bv^{\undisplaced}(\bx) =\summs c_{m,\s}^{\undisplaced}{\bv}_{m,\s}^-(\bx)\,, \label{Eq: undisplacedsingularity}
\end{equation}
where $\summs=\sum_{m=-\infty }^\infty \sum_{\s=0}^1$.

\item 
The above unbounded, active flow is then moved to a coordinate system centered at the domain. The ``shifted'' velocity field depends on the position $\bx$. In the region $|\bx|<|\bxz|$, the velocity field is a superposition of Stokes flow solutions that are non-singular at the origin. In the region $|\bx|>|\bxz|$, the flow consists of Stokes flow solutions that are non-singular at infinity 

\begin{equation}
\bv^{\displaced}(\bx,\bx_0) =\summs  c_{m,\s}^{\displaced,+}(\bxz)\bu_{m,\s}^+(\bx)H(|\bx_0|-|\bx|)+c_{m,\s}^{\displaced,-}(\bxz)\bu_{m,\s}^-(\bx)H(|\bx|-|\bx_0|)
\label{eq: displacedsingularity}
\end{equation}
where $H$ is the step function. 
The coefficients of the transformed flow $c_{ms}^{\displaced,\pm}$  are determined using the transformation procedure found in \citep{greengard1988rapid} and the details are given in \refapp{appendix: FJ transform}.

\item  In the third step, the domain interface is added. The flow inside and outside the domain are 
\begin{equation}
\begin{split}
\bv^{\ins} =   \bv^{\displaced}+\summs  c_{m,\s}^{\ins}\bu_{m,\s}^+,\quad \bv^{\out} =    \summs c_{m,\s}^{\out}\bu_{m,\s}^-.
\end{split}
\label{eq: flows}
\end{equation}
The coefficients for the interior and exterior flows 
are  determined from the boundary conditions.
The continuity of the radial and angular velocity components at the leading order circular boundary, $\bu^\ins|_{r=1} = \bu^\out|_{r=1}$, yields the following relations for the velocity amplitudes
\begin{equation}
\begin{split}
&c_{m,0}^\ins = c_{m,0}^\out-c_{m,0}^{\displaced,-}+|m|(c_{m,0}^\out+c_{m,1}^\out-c_{m,0}^{\displaced,-}-c_{m,1}^{\displaced,-}) \quad |m| > 1,  \\
&c_{m,1}^\ins = c_{m,1}^\out-c_{m,1}^{\displaced,-}-|m|(c_{m,0}^\out+c_{m,1}^\out-c_{m,0}^{\displaced,-}-c_{m,1}^{\displaced,-}) \quad  |m| > 1, \\
&c^\ins_{1,0} = 2(c_{1,0}^\out-c_{1,0}^{\displaced,-}) \quad c_{1,1}^\ins = c_{1,0}^{\displaced,-}-c_{1,0}^\out, \quad |m| = 1, \\
&c_{0,1}^\ins = \frac{1}{2}(c_{0,0}^\out-c_{0,0}^{\displaced,-}), \quad |m| = 0. \\
\end{split} 
\label{eq: continuity}
\end{equation}
The leading order normal component of the hydrodynamic stress evaluated at $r=1$ is written in the form
\begin{equation}
    	 \rhat \cdot \bT = \sum_{m=-\infty}^\infty A^\pm_m \begin{bmatrix}
		c_{m,0}^\pm\\c_{m,1}^\pm
	\end{bmatrix} e^{i m \phi},
\end{equation}
where the matrices $A^\pm_m$ are derived in \refapp{appendix: 2D singularity solutions}. 
The capillary stress depends on the interfacial curvature 
\begin{equation}
  \bnab \cdot\nhat = 1-\epsilon \left(\sum_{m=-\infty}^\infty (1-m^2)f_m e^{i m \phi}\right)+O(\epsilon^2). \label{eq: curvature}
\end{equation}
Accordingly, the leading order normal stress balance  
is 
\begin{equation}
    A_m^-\begin{bmatrix}
    c^\out_{m,0}\\c^\out_{m,1}
\end{bmatrix} -\lambda\left(A_m^-\begin{bmatrix}
c_{m,0}^{\displaced,-}\\c_{m,1}^{\displaced,-}
\end{bmatrix} +A_m^+\begin{bmatrix}
c_{m,0}^\ins\\c_{m,1}^\ins
\end{bmatrix} \right)= - f_m(1-m^2)\rhat.
\label{eq: stress}
\end{equation}
\end{enumerate}
Solving for the velocity amplitudes $c^\out$ and $c^\ins$ using  \refeq{eq: continuity} and  \refeq{eq: stress} yields
\begin{equation}
\begin{split}
     &c^\out_{0,0} = \lambda c^{\displaced,-}_{0,0},\quad  c^\out_{0,1} = c^\out_{1,1} = c^\out_{-1,1} = 0,  \quad c^\out_{\pm1,0} = \frac{2\lambda c_{\pm1,0}^{\displaced,-}}{\lambda+1},\\
	&c^\out_{m,0} = \frac{ -\sgn(m)(|m|-1)if_m+8\lambda c_{m,0}^{\displaced,-}}{4(\lambda+1)}\,,\quad |m| > 1\,,\\
    &
	c^\out_{m,1} = \frac{\sgn(m)(|m|+1)i f_m+8\lambda c_{m,1}^{\displaced,-} }{4(\lambda+1)}\,, \quad |m| > 1,\\ 
\end{split}
\label{eq: coffout}
\end{equation}
\begin{equation}
  \begin{split}
&c^\ins_{0,1} = \frac{\lambda-1}{2} c^{\displaced,-}_{0,0}, \quad c^\ins_{0,0} = 0, \quad c^\ins_{\pm1,0} = 2\frac{\lambda-1}{\lambda+1}c_{\pm1,0}^{\displaced,-}, \quad c^\ins_{\pm1,1} = \frac{1-\lambda}{\lambda+1}c_{\pm1,0}^{\displaced,-},\\
	&c^\ins_{m,0} = \frac{ \sgn(m)(|m|+1)if_m+4(\lambda-1)(c_{m,0}^{\displaced,-}(1+|m|) +c_{m,1}^{\displaced,-}|m|)}{4(\lambda+1)},\quad |m| > 1,\\
	&c^\ins_{m,1} =-\frac{ \sgn(m)(|m|-1)if_m+4(\lambda-1)(c_{m,1}^{\displaced,-}(|m|-1) +c_{m,0}^{\displaced,-}|m|)}{4(\lambda+1)},\quad |m| > 1.\\
\end{split} 
\label{eq: coffin}
\end{equation}

\subsection{Domain interface evolution and steady shape}
 The evolution equation for the interface in the  reference frame co-moving with the domain is given by \refeq{Eq: surface evolution}. To leading order in the capillary number $\epsilon$, the normal vector points in the radial direction, $\nhat = \rhat + O(\epsilon)$, and thus 
\begin{equation}
    \epsilon \frac{\partial f}{\partial t} = \bu|_\surf \cdot \rhat.\\
\end{equation}
For each mode ${|m|\geq 2}$ we have
\begin{equation}
\begin{split}
   &\epsilon \frac{d f_m}{dt}e^{i m \phi} = (c_{m,0}^\out\bu_{m,0}^-|_{r=1}+c_{m,1}^\out\bu_{m,1}^-|_{r=1})\cdot \rhat,\\
& \epsilon \frac{d f_m}{dt} = im (c_{m,0}^\out+c_{m,1}^\out).  
\end{split}
\label{eq: evolution}
\end{equation}
From  \refeq{eq: coffout} and \refeq{eq: evolution},  we obtain the time evolution of the mode coefficients $f_m$, 
\begin{equation}
	 \epsilon \frac{d f_m}{d t} = \frac{-|m|f_m+4i\lambda m (c_{m,0}^{\displaced,-}+c_{m,1}^{\displaced,-})}{2(\lambda+1)}. \label{eq: shapeevolution}
\end{equation}
When the active particle's position is fixed relative to the center of the domain,  the shape relaxes to a stable steady shape, obtained by setting $\frac{d f^\steadyshape_m}{d t}=0$. The steady state  shape mode coefficients are 
\begin{equation}
	f_m^{\steadyshape} = 4 i \sgn(m) \lambda (c_{m,0}^{\displaced,-}+c_{m1}^{\displaced,-}).
    \label{eq: eqmodes}
\end{equation}
Inserting \refeq{eq: eqmodes} into \refeq{eq: coffout} and \refeq{eq: coffin}, the steady state  flow coefficients for ${|m|\geq 2}$ are obtained
\begin{equation}
    \begin{split}
        &c^{\out,\steadyshape}_{m,0} = \lambda\frac{ ( c_{m,0}^{\displaced,-} +c_{m,1}^{\displaced,-})|m|+c_{m,0}^{\displaced,-}-c_{m,1}^{\displaced,-} }{(\lambda+1)},\quad
	c^{\out,\steadyshape}_{m,1} = -\lambda\frac{( c_{m,0}^{\displaced,-}+c_{m,1}^{\displaced,-} )|m|+c_{m,0}^{\displaced,-}- c_{m,1}^{\displaced,-}}{(\lambda+1)},\\ 
    &c^{\ins,\steadyshape}_{m,0}  = -\frac{c_{m,0}^{\displaced,-}+\lambda c_{m,1}^{\displaced,-}+(c_{m,0}^{\displaced,-}+c_{m,1}^{\displaced,-})|m|}{\lambda +1}, \quad c^{\ins,\steadyshape}_{m,1}  = -\frac{c_{m,1}^{\displaced,-}+\lambda c_{m,0}^{\displaced,-}-(c_{m,0}^{\displaced,-}+c_{m,1}^{\displaced,-})|m|}{\lambda +1}. \\ 
    \end{split}
\end{equation}
These coefficients represent a flow which is identical to the flow around a non-deformable, circular domain. This special case is found by setting the normal velocity to  zero at the boundary and enforcing the continuity of the tangential stress (while the normal stress boundary condition is redundant). 

\subsection{Domain motion}
The translational velocity of the domain is defined as {the velocity of its center of mass, which is}
\begin{equation}
\bV_{\domain} = \frac{1}{A} \int_A  \bv^{\ins} dA \label{VolumeAverageVelocity1},
\end{equation}
with $A$ denoting the 2D domain volume. Extra attention is needed to calculate this integral as singularities are presented inside the integration area. Following the approach in \cite{kawakami2025migration}, the domain velocity is expressed as a surface integral 
\begin{equation}
    \bV_{\domain} = \frac{1}{ A} \int_S (\bu_\surf \cdot \nhat)\rhat dS,
\end{equation}
where $S$ is the surface of the domain. At leading order in $\epsilon$, $\nhat=\rhat$, $A=\pi$, and the domain velocity is given by
\begin{align}
\bV_{\domain} =& \sum_{|m|=1} \frac{1}{\pi} \int_0^{2\pi} (c^\out_{m,0}{\bv}_{m,0}^-(1,\phi)\cdot \rhat)\rhat  d\phi
\\
&= c^\out_{-1,0}\begin{bmatrix}
		-i\\-1
	\end{bmatrix}+c^\out_{1,0}\begin{bmatrix}
	i\\-1
	\end{bmatrix} =-4\frac{\lambda}{\lambda+1} \begin{bmatrix}
		\mathcal{I}(c_{1,0}^{\displaced,-})\\\mathcal{R}(c_{1,0}^{\displaced,-})
	\end{bmatrix}.
       \label{eq: domainvelocity}
\end{align}
Only the coefficients for the exterior flow, $c^\out_{m,\s}$, with $(|m|=1,\s=0)$ have a non-zero contribution to the domain velocity as shown in the surface integral calculation in \refeq{eq: intflow}. An important consequence from the transform in \refapp{appendix: FJ transform} is that only a rotlet $(m=0,\s=0)$, source dipole $(|m|=1,\s=0)$ and force dipole $(|m|=1,\s=1)$ can excite domain translation as these are the only singularities with non-zero $c^{\undisplaced,-}_{\pm1,0}$ coefficients when transformed. In addition, the domain velocity at steady state shape is identical to the velocity of the circular domain.

\subsection{Trajectory of the active particle}
The active particle inside the domain is advected by the flow (generated by the singularity and  reflected by the boundary). The orientation of the active particle changes proportional to the cross product between the vorticity and $\phat$. In the reference frame co-moving with the domain, 
\begin{equation}
\begin{split}
    &\frac{d \bx_0}{dt} = {V}_\particle \bm{\hat{p}}-\bV_{\domain}+\summs  c_{m,\sigma}^{\ins}\bu_{m,\sigma}^+,\\ 
 &\frac{d\phat}{dt} = \frac{1}{2}\left( \sum_{m=-\infty }^\infty c_{m,1}^{\ins}\bnab  \times\bu_{m,1}^+\right)\times \bm{\hat{p}}.
\end{split}
\label{eq: particleev1}   
\end{equation}
Flows with $\s=0$ have no contribution in the orientation dynamics as they are irrotational. In terms of the orientation angle $\alpha = \cos^{-1}(\phat \cdot \xhat)$, \refeq{eq: particleev1}, using \refeq{eq: flowcurl}, can be rewritten as
\begin{equation}
     \frac{d\alpha}{dt}=\sum_{m=-\infty}^{\infty} -2c_{m,1}^{in}r_0^{|m|}e^{i m \phi_0} 
     (|m|+1)
     \label{eq: particleev3} .  
\end{equation}
\section{Results and discussion}
\label{sec: Results}

\subsection{Stokeslet}
{It is instructive to consider the case of the point force at the domain center and relate it to the works of \cite{saffman1976brownian,yariv2025mobility,schutte2017size}. 
To remedy the Stokes paradox, the membrane is assumed to be of finite size and  confined by a circular boundary at $r=L$, where no-slip conditions are imposed. 
In \refeq{eq: StokesEqinside}, the Stokeslet is represented by $\textbf{f}(\bx)=-\xhat\delta(\bx) $.
The flow it generates is given by
\begin{align}
& \bu_{stokes}=  \bu_{stokes,1} +  \bu_{stokes,-1}, \\
& \bu_{stokes,\pm 1} =
\frac{1}{8\pi\lambda}e^{\pm i\phi}\left((1-\ln(r))\rhat  -i\ln(r)\phihat \right).
\end{align}
The velocity boundary conditions are
\begin{equation}
\bv^{\ins} = \bv^{\out} \,\, \text{at }r=1, \quad  \bv^{\out}=\mathbf{0} \,\, \text{at }r=\frac{L}{\Rdomain}  \label{BC: continuity2}.
\end{equation}
To satisfy these boundary conditions, homogeneous solutions of the Stokes equations are added to the singular stokeslet flow. The homogeneous solution inside the domain must be regular at the origin. Because the outer fluid is bounded, homogeneous solutions that diverge at infinity may also be included in the outer region. Furthermore, the symmetry of the stokeslet restricts the homogeneous contributions to the $|m|=1$ modes. For simplicity, only the $m=1$ modes are considered explicitly, since the $m=-1$ modes are their complex conjugates. The total physical flow is therefore obtained as twice the real part of the $m=1$ contribution.

Accordingly, we assume
\begin{equation}
\bu^{\ins} =   c_{1,0}^{\ins}\bu_{1,0}^++c_{1,1}^{\ins}\bu_{1,1}^+ +  \bu_{stokes, 1} ,\quad \bu^{\out} =  c_{1,0}^{\out,-}\bu_{1,0}^- +  \lambda \bu_{stokes,1}+c_{1,0}^{\out,+}\bu_{1,0}^++c_{1,1}^{\out,+}\bu_{1,1}^+.
\label{eq: stokesletflows}
\end{equation}
The amplitude of the stokeslet contribution in the outer flow is fixed by requiring the net force transmitted across the interface to be continuous. Equivalently, the net force exerted by the inner domain on the outer fluid must be equal and opposite to the force exerted by the outer fluid on the domain at $r=1$, i.e.
\begin{equation}
\int_S -\rhat \cdot \bT^\ins ds = - \int_S \rhat \cdot \bT^\out  ds.
\end{equation}
Since the stokeslet is the only term that produces a nonzero net force, this condition uniquely determines the outer stokeslet amplitude.

In addition, the tangential stress must be continuous across the interface at $r=1$.
For $m=1$, the tangential stress condition can be written using the matrices $A^\pm_1$ derived in \refapp{appendix: 2D singularity solutions} as
\begin{align*} \rhat \cdot (\bT^\out- \bT^\ins)\cdot \phihat=\left( A_1^-\begin{bmatrix} c^{\out,-}_{1,0}\\0 \end{bmatrix}+A_1^+\begin{bmatrix} c_{1,0}^{\out,+}\\c_{1,1}^{\out,+} \end{bmatrix} -\lambda\left (A_1^+\begin{bmatrix} c_{1,0}^\ins\\c_{1,1}^\ins \end{bmatrix} \right)\right)\cdot \phihat=0 . \end{align*}

Here, the stress contributions from the stokeslet terms cancel.
Together with the velocity boundary conditions, this yields a system of five equations for the five unknown coefficients. After solving for $\bu^\out$, the domain velocity is obtained by integrating the normal component of the outer velocity over the domain boundary. Since only the $m=1$ mode is considered explicitly, the result is twice its real part

\begin{flalign}
& \bV_{d} = 2Re\left(\frac{1}{ \pi} \int_0^{2\pi} (\bu^\out|_{r=1}\cdot \rhat)\rhat d\phi \right)=\frac{1}{4\pi}\left(\log (\omega)-\frac{\left(\omega^2-1\right)
\left(\left(\omega^2-1\right) \lambda
+1\right)}{\left(\omega^4-1\right) \lambda
+\omega^4+1}\right)\xhat, \quad \omega \equiv \frac{L}{\Rdomain}.
\end{flalign}
In dimensional form, the drop velocity becomes
\begin{flalign}
\bV_{d} =\frac{F}{4 \pi \eta_m}\left(\ln(\omega)-\frac{\left(\omega^2-1\right)
\left(\left(\omega^2-1\right) \lambda
+1\right)}{\left(\omega^4-1\right) \lambda
+\omega^4+1}\right)\xhat.
\label{eq: fluidmobility}
\end{flalign}
Finally, taking the limit $\lambda \to \infty$ recovers the result for a concentric rigid particle, with $L=2\ell_\sd$ identified with the Saffman-Delbrück  length
\begin{flalign}
\bV_{d} =\frac{F}{4 \pi \eta_m}\left(\ln (\omega)-\frac{
\omega^2-1
}{\omega^2+1
}\right)\xhat.
\label{eq: solid}
\end{flalign}
Equation \refeq{eq: solid} recovers the mobility used in Eq. 3 of \cite{schutte2017size} to determine the diffusivity of a solid domain. By contrast, using the mobility given in \refeq{eq: fluidmobility} yields a modified mobility coefficient appropriate for a fluid domain, as is the case for lipid rafts. The non-concentric case of the solid domain and outer boundary is later derived in \cite{yariv2025mobility}. This equation differs from that derived in \cite{saffman1976brownian}, where the coefficient $c_{1,1}^{\out,+}$ in \refeq{eq: stokesletflows} is set to zero. As a result, the full no-slip condition cannot be imposed at the outer boundary, and only the radial velocity component is set to zero.}
\subsection{A stresslet}
\begin{figure}
\centering
\includegraphics[width = \textwidth]{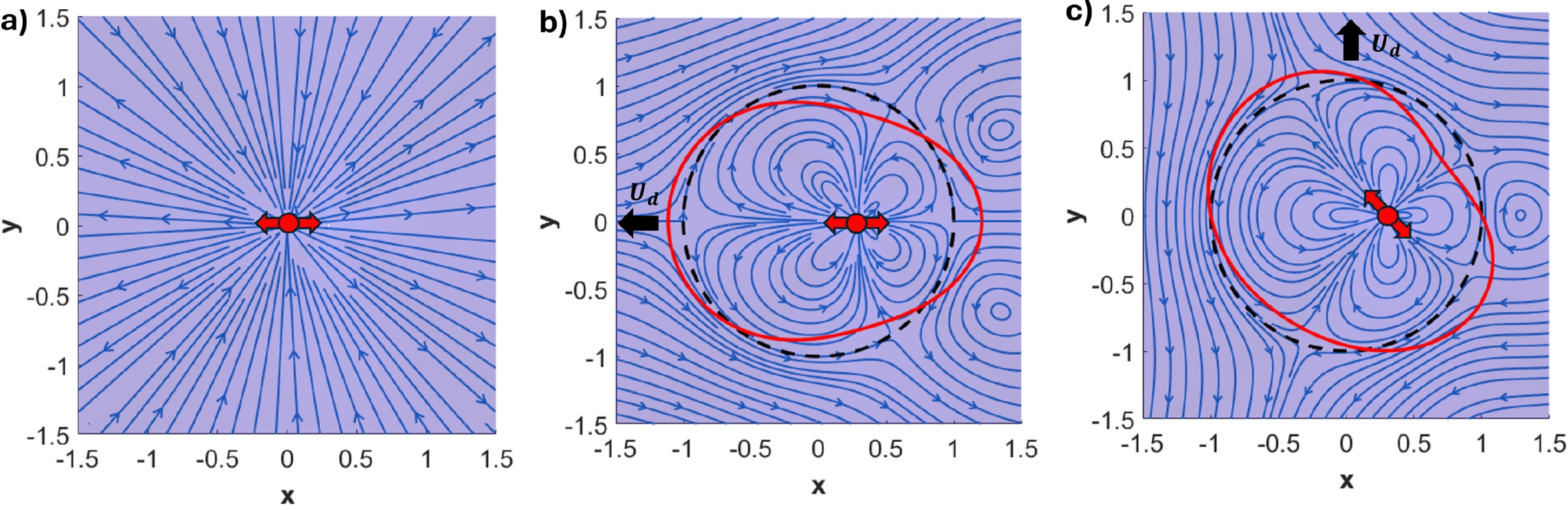}
\caption{\textbf{Flow of a stresslet in free space and confined inside a circular domain.} (a) Flow about an unconfined stresslet placed at the origin. (b), (c) Flow created by a stresslet confined in  a domain and  located off-center at $(r_0,\phi_0) =(0.3,0)$. The stresslet is tilted relative to the line connecting its position to the domain’s center, with angles $\alpha = 0$ and $\alpha = 3\pi/4$, respectively. In all plots $Ca=0.04$ and $\lambda=1$. The red dot represents the location of the stresslet. The black line represents the undisturbed domain shape and the red line represents the equilibrium shape of the domain. The black arrow represents the direction of motion of the domain. For (b) it points  to the left and for (c) it points upwards. 
}
\label{fig: stressletflow}
\end{figure}

A stresslet flow is described by \citep{lushi2014fluid} $\textbf{f}(\bx) =\frac{1}{\lambda}\bnab \cdot\left( 4\pi( \phat\phat) \delta(\bx - \bf{x_0})\right)$ and $g(\bx)=0$ in \refeq{eq: StokesEqinside}, with $\phat$ oriented at an angle $\alpha$ relative to the x-axis. The expression for the unbounded  stresslet flow in a coordinate system centered at the particle is given by
\begin{equation}
    \bv^{\undisplaced}(\bx) = \frac{1}{\lambda r}(2(\phat\cdot \rhat)^2\rhat-\rhat)=\frac{1}{\lambda}\frac{\cos(2(\phi-\alpha))}{r}\rhat.
    \label{eq: closestresslet}
\end{equation}
This flow  is illustrated in \reffig{fig: stressletflow}(a).The flow coefficients
in the velocity representation \refeq{Eq: undisplacedsingularity} are 
\begin{equation}
     c_{-2,1}^{\undisplaced}=\frac{1}{2\lambda}e^{i(2\alpha+\frac{\pi}{2})},\quad  c_{2,1}^{\undisplaced}=\frac{1}{2\lambda}e^{-i(2\alpha+\frac{\pi}{2})}.
\end{equation}
The domain  velocity is obtained from  \refeq{eq: domainvelocity} using the coefficients \refeq{eq: coffstressimag} yielding at steady-state
\begin{equation}
\begin{split}
    \bV_{\domain}=&-2r_0\frac{1}{\lambda+1} \begin{bmatrix}
		\cos(2\alpha-\phi_0)\\ 	\sin(2\alpha-\phi_0)
	\end{bmatrix}=2r_0\frac{1}{\lambda+1}(\rhat-2(\phat\cdot \rhat) \phat)\\
    &= -2r_0\frac{1}{\lambda+1}\left(\cos(2(\alpha-\phi_0))\rhat+\sin(2(\alpha-\phi_0))\mathbf{\hat{\phi}}\right) .
    \label{eq: domainvelstress}
    \end{split}
\end{equation}
The domain velocity induced by the stresslet depends on the particle’s position, $\mathbf{x}_0 = (r_0,\phi_0)$, and includes a component misaligned with the active particle's orientation. In the outer expansion about the origin, a displaced stresslet generates a $(|m|=1, s=0)$ irrotational mode that drives translation. This mode corresponds to a source dipole at the origin. Its orientation depends on the particle location and inherits the two-fold symmetry of the stresslet. As a result, the corresponding multipole dipole flow is generally not aligned with the stresslet orientation.

The flow patterns for the confined stresslet are shown in
in \reffig{fig: stressletflow}(b) and \reffig{fig: stressletflow}(c), where the stresslet is held fixed relative to the domain center, while the domain shape is allowed to evolve and reach a steady state. The domain translates with velocity $\bV_{\domain}$ indicated by the black arrow. The direction will depend on $\alpha$ and $\phi_0$ and the magnitude on the radial location $r_0$ of the stresslet. 
In \reffig{fig: stressletflow}(b), the stresslet is located at $(x, y) = (0.3, 0)$, with orientation angle $\alpha = 0$, and the domain moves to the left.  
The red contour represents the steady state deformed shape of the domain, while the dashed black line indicates its initial circular shape. The domain bulges in regions where the stresslet driven flow is directed outward and indents where the flow is inward. The mismatch between interior and exterior flows at the domain interface, nonzero normal velocity, is a characteristic of the leading-order asymptotic solution.   In \reffig{fig: stressletflow}(c), the stresslet remains at the same location, but is reoriented at an angle $\alpha = 3\pi/4$ with  domain velocity $\bV_{\domain}$ pointing upwards. For non-zero $\phi_0$, the domain velocity will generally be misaligned with the stresslet direction. In \reffig{fig: stressletflow}, the evaluation of the active flow is done using \refeq{eq: closestresslet} to avoid issues with the convergence of the transformed solution at $|\bx|=|\bxz|$. The confinement-induced flows in \refeq{eq: flows} are evaluated using the steady shape coefficients in \refeq{eq: coffstresseq}.

The dynamics of a motile stresslet in the domain reference frame are given by \refeq{eq: particleev1}. For the scenarios of a non-deformable domain or a quasi-steady state where $\frac{1}{\epsilon} \gg V_\particle$, the steady state coefficients are given by \refeq{eq: coffstresseq}. 
Inserting the steady-shape coefficients from \refeq{eq: coffstresseq} and the domain velocity from \refeq{eq: domainvelstress} into \refeq{eq: particleev1} yields 
\begin{equation}
\begin{split}
\frac{d r_0}{dt} =& {V}_\particle\cos(\alpha-\phi_0) +\frac{2r_0\cos(2(\alpha-\phi_0))}{1+\lambda}+\sum_{m=-\infty}^\infty\left( c_{m,0}^{\ins,\stresslet,\steadyshape}\bu_{m,0}^+(\bxz)\cdot \rhat+c_{m,1}^{\ins,\stresslet,\steadyshape}\bu_{m,1}^+(\bxz)\cdot \rhat\right), \\
r_0\frac{d \phi_0}{dt} =&  {V}_\particle\sin(\alpha-\phi_0) +\frac{2r_0\sin(2(\alpha-\phi_0))}{1+\lambda}+\sum_{m=-\infty}^\infty\left( c_{m,0}^{\ins,\stresslet,\steadyshape}\bu_{m,0}^+(\bxz)\cdot \bm{\phihat}+c_{m,1}^{\ins,\stresslet,\steadyshape}\bu_{m,1}^+(\bxz)\cdot \bm{\phihat}\right), \\
\frac{d\alpha}{dt}=&\sum_{m=-\infty,m\neq 1,0}^{\infty} -2\frac{r_0^{| m| -2} \left((1-r_0^2)| m| +\lambda  r_0^2-1\right)e^{-i \sgn(m) \left(2\alpha+(| m|-2)\phi_0 +\frac{\pi }{2}\right)}}{2\lambda(\lambda +1)}r_0^{|m|}e^{i m \phi_0} 
     \left(|m|+1\right)\\
     &+ 4\sin(2(\alpha-\phi_0))\frac{\lambda-1}{(1+\lambda)\lambda}r^2_0.\\
\end{split}
\label{eq: sum}
\end{equation}

 The sums in \refeq{eq: sum} are a geometric series of $r_0$ and when summed result in

\begin{equation}
    \begin{split}
&\frac{d r_0}{dt} =V_p\cos(\alpha-\phi_0)+\frac{2 r_0\cos (2(\alpha-\phi_0))}{
\lambda\left(r_0^2-1\right)},\\
& \frac{d \phi_0}{dt} = \frac{V_p}{r_0}\sin(\alpha-\phi_0)+\frac{2\left(1+\lambda(1- r_0^2)\right) \sin (2 (\alpha-\phi_0))}{\lambda(\lambda +1)
\left(r_0^2-1\right)},\\
& \frac{d \alpha}{dt} = -\frac{2r^2_{0}(-1+\lambda(-2+r_0^2))\sin(2(\alpha-\phi_0))}{\lambda(r_0^2-1)^2(1+\lambda)}.
    \end{split} 
    \label{eq: stresseqs}
\end{equation}
This coupled system of differential equations describes the motion of an active stresslet within the domain.
When the self-propulsion $V_p$ term is non-zero, the equations of motion have fixed points if $\alpha - \phi_0 = 0$, i.e., when the stresslet points radially outward, as discussed in \refsec{sec: jacobianStress}. The fixed points are given by $(r_0, \phi_0, \alpha_0) = (r_c, \phi_c, \phi_c)$, where the critical radius is
$r_c = -1 + \sqrt{1 + (\lambda V_p)^{-2}} $ with $0 < r_c < 1$. As shown in \refsec{sec: jacobianStress}, this fixed point corresponds to a saddle. In the case the self-propulsion term $V_p$ is zero, there is only an unstable saddle fixed point at $r_c=0$. 
As a result, a self-propelled stresslet simply ``swims" to the boundary. 
The advection by the self-generated flow is $\sim 2(\alpha-\phi_0)$, while the self-propulsion term has a different symmetry,  $\sim(\alpha-\phi_0)$. Thus, in the radial motion \refeq{eq: stresseqs} the advection can not counteract the self-propulsion. Additional modeling constraints must be introduced to prevent the stresslet from leaving the domain.

\subsection{Source dipole}
\begin{figure}
\begin{center}
{\includegraphics[width=\textwidth]{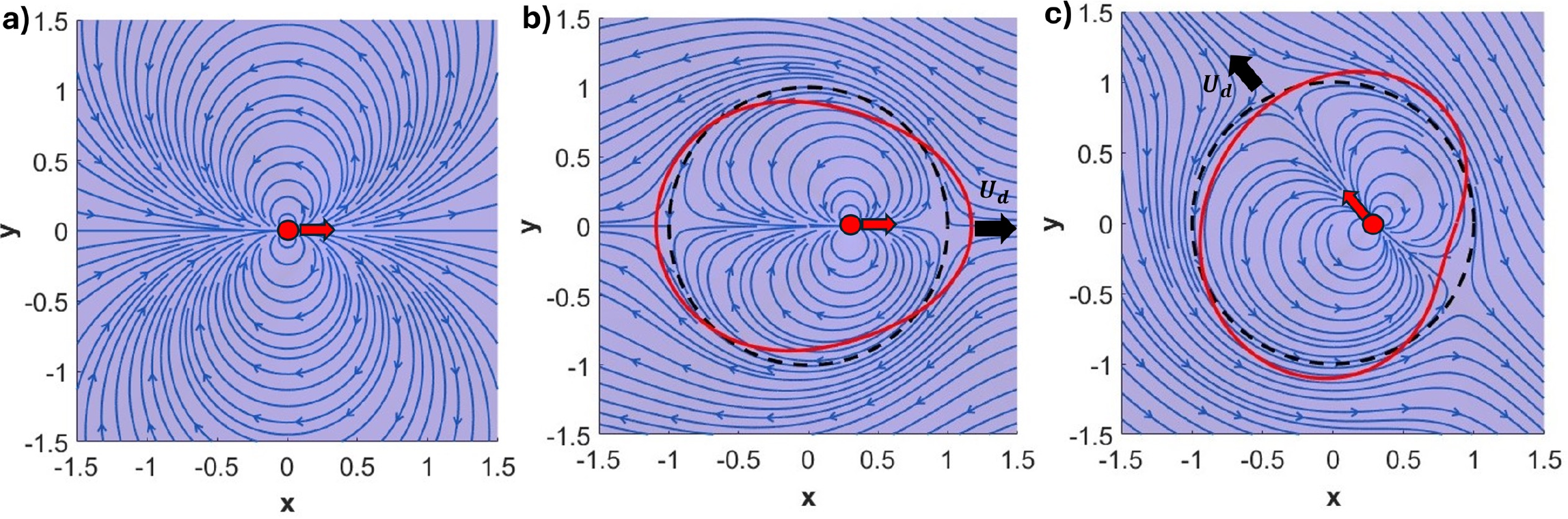}}
\end{center}
\caption{\textbf{Flow due to a source dipole in free space and confined inside a circular domain.} (a) Flow about an unconfined source dipole at the origin; all streamlines are closed. (b), (c) Flow created by a source dipole confined in  a circular  domain and  located off-center at $(x,y)= (0.3,0)$.   The dipole is tilted relative to the line connecting the dipole and the center of mass of the domain at angle  $\alpha =0$ and   $\alpha =3\pi/4$ correspondingly. The streamlines outside the domain are open, indicating domain motion. In all  plots $Ca=0.05$ and $\lambda=1$. The red dot represents the location of the dipole. The black line represents the undisturbed domain shape and the red line represents the equilibrium shape of the domain. The black arrow represents the migration direction of the domain; the fluid is being pumped in the opposite direction. 
}
\label{fig: dipoleflow}
\end{figure}

The far field of an idealized 2D neutral squirmer is a dipolar flow,  \citep{alas2021swimming,ahana2019confinement,matas2014hydrodynamic}. 
A source dipole corresponds to $\textbf{f}(\bx)=0$ and $g(\bx) = - 4\pi\phat \cdot \bnab \delta(\bx - \bf{x_0})$ in \refeq{eq: StokesEqinside} where $\phat=[\cos{(\alpha)},\sin{(\alpha)}]^{T}$.
 The expression for the unbounded source dipole flow in a coordinate system centered at the particle is given by
 \begin{equation}
    \bv^{\undisplaced}(r,\phi) = 2\frac{(2\rhat\rhat - \bm{I})\cdot\phat}{r^2}=\frac{2}{r^2}\left(\cos(\phi-\alpha)\rhat+\sin(\phi-\alpha)\phihat\right).
    \label{eq: closedip}
\end{equation}
This flow in unbounded space is illustrated in \reffig{fig: dipoleflow}(a). The flow coefficients
in the velocity representation \refeq{Eq: undisplacedsingularity} are 
\begin{equation}
  c_{-1,0}^{\undisplaced}=e^{i(\alpha+\frac{\pi}{2})},\quad   c_{1,0}^{\undisplaced}=e^{-i(\alpha+\frac{\pi}{2})}.
\end{equation}
To calculate the domain velocity, the expression in \refeq{eq: domainvelocity} is evaluated using the coefficients in \refeq{eq: coffdip} yielding at steady state
\begin{equation}
    \bV_{\domain}=4\frac{\lambda}{\lambda+1} \begin{bmatrix}
		\cos(\alpha)\\ 	\sin(\alpha)
	\end{bmatrix} = 4\frac{\lambda}{\lambda+1} \phat =  4\frac{\lambda}{\lambda+1}\left(\cos(\alpha-\phi_0)\rhat+\sin(\alpha-\phi_0)\mathbf{\hat{\phi}}\right).
    \label{eq: domainveldip}
\end{equation}
Unlike the domain velocity induced by a stresslet \refeq{eq: domainvelstress}, the velocity induced by a source dipole \refeq{eq: domainveldip} is independent of the dipole position and is always aligned with the dipole orientation $\phat$. In the outer expansion about the origin, a displaced dipole generates a hierarchy of multipole modes, but only the $(|m|=1, \s=0)$ mode contributes to translation. This mode corresponds to a source dipole at the origin with the same orientation and strength as the original dipole. The flow generated by this multipole drives the drop in the direction it points. All other modes have zero net contribution to the domain velocity. Therefore the drop moves along $\hat{\mathbf p}$ regardless of the dipole location.

The flow patterns for the confined  dipole are shown
in \reffig{fig: dipoleflow}(b) and \reffig{fig: dipoleflow}(c), where the dipole is held fixed relative to the domain center, while the domain shape is allowed to evolve and reach a steady state.  
In both cases, the domain translates with velocity $\mathbf{U_d}$, in the same direction as the dipole orientation.  
In \reffig{fig: dipoleflow}(b), the dipole direction is colinear with the domain center and the flow is axisymmetric.
The red contour represents the steady state deformed shape of the domain, while the dashed black line indicates its initial circular shape.  
The domain bulges in regions where the dipole driven flow is directed outward and indents where the flow is inward. Note that the velocity is continuous across the circular contour, not the deformed interface.  
In \reffig{fig: dipoleflow}(c), the dipole remains at the same location, but is reoriented with $\alpha = 3\pi/4$.
In \reffig{fig: dipoleflow}, the evaluation of the active flow is done using \refeq{eq: closedip} to avoid issues with the convergence of the transformed solution at $|\bx|=|\bxz|$. The confinement-induced flows in \refeq{eq: flows} are evaluated using the steady shape coefficients in \refeq{eq: coffdipeq}.

Interestingly, domain deformations are different for the stresslet and the dipole, \reffig{fig: stressletflow}(b) and \reffig{fig: dipoleflow}(b), despite both cases producing domain translation in the same direction. For a source dipole, the interface bulges outward in the direction of the domain velocity $\bV_{\domain}$, whereas for a stresslet the deformation is reversed. 
In our linear perturbation analysis, shape deformation and  domain translation are decoupled. 
In the small-deformation regime $\mathrm{Ca}\ll 1$ and a steady state shape, the domain velocity is the same as for a circular domain and hence independent of the interfacial deformation. 
Mathematically, translation is related to $(|m|=1,\s=0)$, while deformation is determined by modes $|m|\ge 2$, and within the linear perturbation analysis these modes do not couple.

\subsubsection{Dynamics of a motile singularity in a circular domain: periodic trajectories}
\label{sec: dipoledyn}

The motion of the dipole in the domain reference frame is given by \refeq{eq: particleev1}. For the scenarios of a non-deformable domain or a quasi-steady state where $\frac{1}{\epsilon} \gg V_\particle$, the steady state shape and flow are given by the coefficients \refeq{eq: coffdipeq}. Substituting the inner steady-shape coefficients from \refeq{eq: coffdipeq} and the domain velocity from \refeq{eq: domainveldip} into \refeq{eq: particleev1} yields an infinite series analogous to that obtained for the stresslet \refeq{eq: sum}. These terms form a geometric series in $r_0$, which, when summed, result in

\begin{flalign}
    &\frac{d r_0}{dt} =\left(V_p-2\frac{2+(1+\lambda)r_0^2}{(1+\lambda)(r_0^2-1)^2}\right)\cos(\alpha-\phi_0) \label{eq: radeq1},\\
    &\frac{d \phi_0}{dt} =\frac{1}{r_0}\left(V_p +2\frac{(-2+(1-3\lambda)r_0^2+2
    \lambda r_0^4) }{(1+\lambda)(r_0^2-1)^2}\right)\sin(\alpha-\phi_0)
        \label{eq: angulareq},\\
    &\frac{d\alpha}{dt}= \left(-4\frac{ r_0 (-2 + \lambda (2 - 3 r_0^2 + r_0^4))}{(1 + \lambda) (1-r_0^2)^3}\right)\sin(\alpha - \phi_0).
    \label{eq: dipoleeq}
\end{flalign}
This coupled system of differential equations describes the hydrodynamic motion of an active dipole within a domain.
If  $V_p>\frac{4}{1+\lambda}$, there is special location where the particle radial velocity vanishes (see \refapp{sec: proofF}); the location is 
\begin{equation}
    r_{c} = \sqrt{\frac{(1+\lambda)(1+V_p)-\sqrt{(1+\lambda)(1+6V_p+\lambda(1+2V_p))}}{V_p(1+\lambda)} }.
    \label{eq: rc}
\end{equation}

\begin{figure}
\centering
\includegraphics[width = \textwidth]{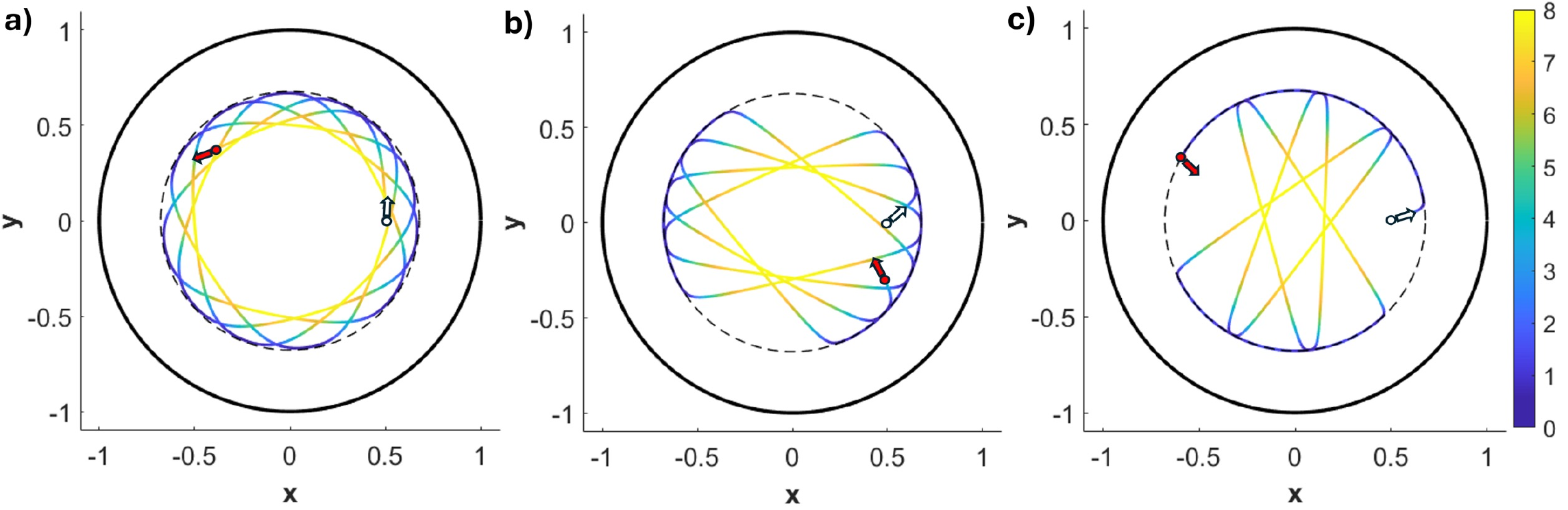}
\caption{
\textbf{Periodic motion of a motile dipole inside a circular  domain:}
 Dipole initially positioned inside at $(r_0,\phi_0) = (0.5,0)$ with initial orientation (a) $\alpha= \frac{\pi}{2}$, (b)  $\alpha= \frac{\pi}{5}$, and (c) $\alpha= \frac{\pi}{10}$. The color bar  represents the dimensionless speed of the singularity. The dashed line represents the radial fixed point given by \refeq{eq: rc}, in this case $r_c=0.68$. The white arrow represents the initial position and the red arrow represents the position at the end of the computation. The propulsion velocity of the dipole is $V_p=10$ . The simulations are run for 5 units of dimensionless time. 
$\lambda=1$. }
\label{fig: analyticsource}
\end{figure}

If the dipole's initial location is $r_0>r_c$, it is 
moves monotonically 
toward the domain boundary. However, if the dipole's initial location is $r_0<r_c$, its motion becomes more intricate. It swims and reorients in such a way that its trajectory always remains within the circle of radius $r_c$, 
undergoing periodic motion. While, the unbounded swimmer is persistent, confinement induces periodic, orbiting motions restricted within $r<r_c$. \reffig{fig: analyticsource} illustrates the effect of the dipole’s initial location and orientation on its trajectory. In free space, the dipole’s flow is irrotational. When the dipole is confined inside a domain, the flow must satisfy the velocity continuity condition \refeq{eq: continuity} and the stress condition \refeq{eq: stress}, which introduce flow components that are not irrotational. Therefore, while an unbounded dipole does not produce flow capable of rotating other singularities, the confinement-induced flows can.

For $r_0<r_c$ in the radial \refeq{eq: radeq1} and in the regime $V_p>\frac{4}{1+\lambda}$, the radial velocity is positive when $-\frac{\pi}{2}<\alpha-\phi_0<\frac{\pi}{2}$  and negative when  $\frac{\pi}{2}<\alpha-\phi_0<\frac{3\pi}{2}$. Therefore when the dipole orientation $\alpha$ is anti-aligned with the polar coordinate $\phi_0$ it  swims towards the center of the domain and when it is aligned with the polar coordinate it swims outwards. In the dipole reorientation \refeq{eq: dipoleeq}, the magnitude of angle change will be largest as $r_0$ approaches 1. For the case $\lambda\leq 1$, the dipole will turn  anticlockwise when $0<\alpha-\phi_0<\pi$ and clockwise when $\pi<\alpha-\phi_0<2\pi$. For other values of $\lambda>1$ the turning direction will depend on $r_0$.

In \reffig{fig: analyticsource}, trajectories of a motile dipole governed by \refeq{eq: dipoleeq} inside a steady shape domain are plotted for several initial conditions.  
As we are primarily interested in the regime where domain deformations remain small, the dipoles are initialized inside the critical radius $r_c$ defined in \refeq{eq: rc}.  
Dipoles cannot cross this boundary and as they approach radius $r_c=0.68$, it decelerates and reorientation becomes more significant.    
This behavior is illustrated by the color scale, which indicates faster dipole motion at smaller radial positions. 
The trajectories are obtained by numerically integrating the equations of motion using a Runge–Kutta 45 method.  
The simulations also show that even with the same initial position, different initial dipole orientations can lead to a wide variety of trajectories. Additionally, we observe that larger initial dipole orientations lead to trajectories that remain more tightly confined near the limiting radius $r_c$.

\begin{figure}
\centering
\includegraphics[width = 0.75\textwidth]{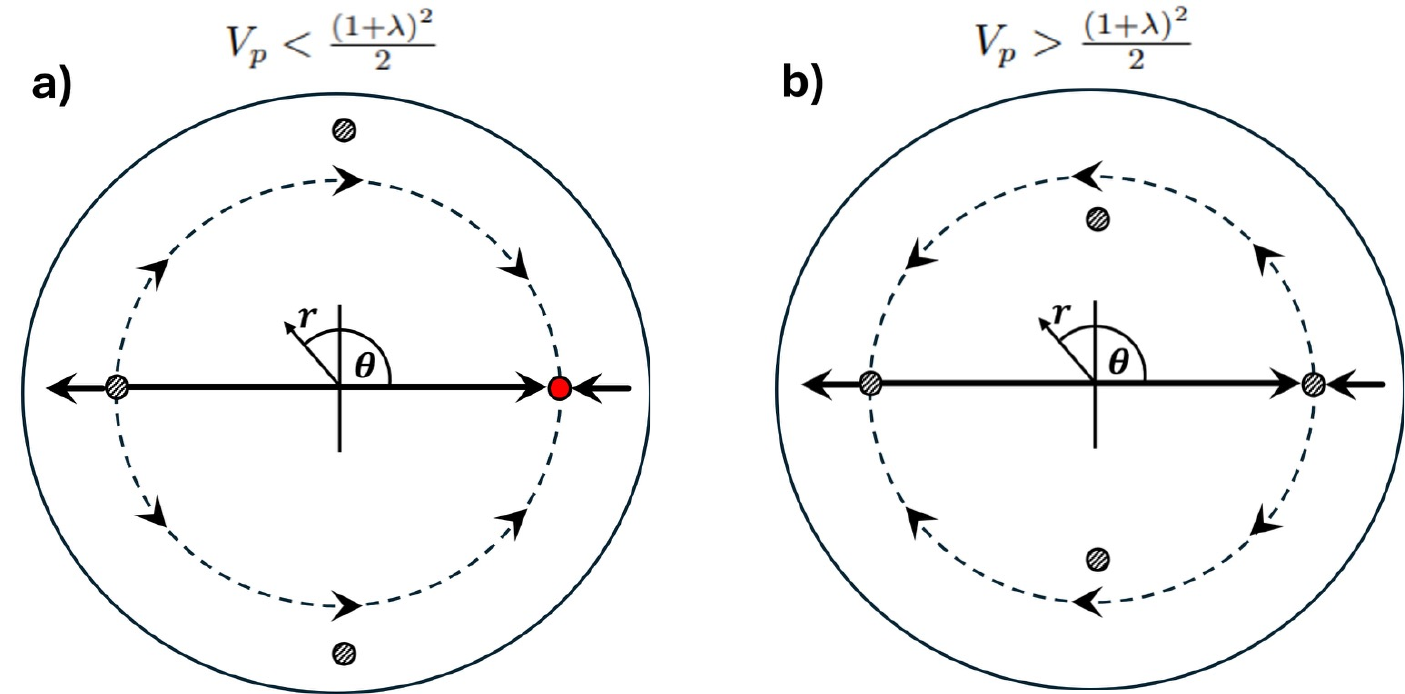}
\caption{{\bf{Two regimes for the dynamical system \refeq{eq: polarred}}}. The dashed line represents the radius $r_c$ where the radial particle velocity is zero. Two fixed points are found on this circle. One unstable at $\theta = \pi$ and one stable at $\theta=0$ when ${V_p<{(1+\lambda)^2}/{2}}$ and both unstable when $V_p>{(1+\lambda)^2}/{2}$ . The other two fixed-points are at radius $r_p$, where $r_p$ is the zero of the function $H(r_p)$ in the angular evolution ODE. The parameters controlling stability are particle swim velocity $V_p$ and domain viscosity ratio $\lambda$. The fixed-points at $r_c$ are saddles when  ${V_p<{(1+\lambda)^2}/{2}}$  and centers with closed orbits around them when  ${V_p>{(1+\lambda)^2}/{2}}$.}
\label{fig: stability} 
\end{figure}

Next, we analyze the particle dynamics in detail. 
The system of equations formed by \refeq{eq: radeq1}, \refeq{eq: angulareq} and \refeq{eq: dipoleeq} can be reduced to a two equation system in terms of the particle radial position $r_0$ and modified angle $\theta_0 = \alpha - \phi_0$
\begin{equation}
\begin{split}
&\frac{d r_0}{dt}= F(r_0)\cos(\theta_0),\quad F(r_0)\equiv\left(V_p-2\frac{2+(1+\lambda)r_0^2}{(1+\lambda)(r_0^2-1)^2}\right),\\
&\frac{d\theta_0}{dt}= H(r_0)\sin(\theta_0), \quad  H(r_0)\equiv \left(\frac{4-2(-1+\lambda)r_0^2+2(1+\lambda)r_0^4}{r_0(1+\lambda)(1-r_0^2)^3}-\frac{1}{r_0}V_p\right).
\end{split}
\label{eq: polarred} 
\end{equation}
There are four fixed points in the system. 
Two are located at $(r_0,\theta_0) = (r_c,\theta_c)$ with $\theta_c = 0$ or $\pi$, and where $r_c$ denotes the unique real solution of  $F(r_c) = 0$ in the range $0<r_0<1$, given by \refeq{eq: rc} and derived in \refsec{sec: proofF}.
The other two fixed points are at $(r_p,\theta_p)$ with $\theta_p = \frac{\pi}{2}$ or $\frac{3\pi}{2}$, where $r_p$ is the unique positive solution of $H(r_p) = 0$ found in $0<r_0<1$, discussed in \refsec{sec: proofH}. The fixed point at $(r_c,0)$ corresponds to a dipole oriented radially outward that remains stationary, while the fixed point at $(r_c,\pi)$ corresponds to a dipole oriented radially inward that also remains stationary. 
At radius $r_p$, the fixed point at $(r_p,  \frac{\pi}{2})$ corresponds to a dipole orbiting counterclockwise inside the domain at radius $r_p$, with the dipole making an angle of  $\frac{\pi}{2}$ with the radial vector. The fixed point at $(r_p, \frac{3\pi}{2})$ corresponds to motion in the clockwise direction, with the dipole making an angle of  $\frac{3\pi}{2}$ with the radial vector.
\\
The stability of the fixed points at $r_c$ depends on the relative magnitude of $r_c$ and $r_p$.
If $r_c<r_p$, corresponding to the condition $V_p<(1+\lambda)^2/2$, the fixed point at $(r_c,0)$ is stable, while the fixed point at $(r_c,\pi) $ is unstable. Furthermore, the fixed points $(r_p,\theta_p)$ are saddles. 
If $r_c>r_p$, which corresponds to the condition $V_p>(1+\lambda)^2/2$, both fixed points $(r_c,0)$ and $(r_c,\pi)$ are unstable. The fixed points $(r_p,\theta_p)$ are centers. This behavior is summarized in \reffig{fig: stability}. In addition, the system has a separatrix along the mid-horizontal line, which separates the phase space into two distinct regions. This corresponds to the fact that the dipole can either move clockwise or counterclockwise around the domain but cannot switch between the two. The stability analysis is shown in detail in \refsec{sec: jacobian}. 
One interesting regime is $\lambda \leq 1$, together with the restriction $V_p\geq\frac{4}{1+\lambda}$, which implies $V_p>(1+\lambda)^2/2$. This results in no stable fixed points and only close orbit cycles inside the region $r_0<r_c$, as seen in \reffig{fig: analyticsource}.
\\

In the case $V_p>(1+\lambda)^2/2$, the fixed points at $(r_p,\theta_p)$ are centers. By integrating the separable governing equations $\frac{d\theta_0}{dr_0}=\frac{H(r_0)}{F(r_0)}\tan(\theta_0)$,  from initial condition $(r_i,\theta_i)$ with $r_i<r_c$, an implicit closed curve can be found 
\begin{equation}
    |\sin{\theta_0}|e^
    {-G(r_0)}=|\sin(\theta_i)|e^
    {-G(r_i)},
    \label{eq: impcurve}
\end{equation}
with $G(r_0)$ given by 
\begin{equation}
    \begin{split}
  &G(r_0)=-\frac{\left(\lambda ^2+f(\lambda)\sqrt{\lambda +1}+3 \lambda +2\right) \log \left(\sqrt{\lambda +1} \left(\left(r_0^2-1\right) V_p-1\right)-f(\lambda)\right)}{\sqrt{\lambda +1} (\lambda +3) f(\lambda)} \\
   &+\frac{\left(\lambda ^2-\sqrt{\lambda +1} f(\lambda)+3
   \lambda +2\right) \log \left(\sqrt{\lambda +1} \left(\left(r_0^2-1\right) V_p-1\right)+f(\lambda)\right)}{\sqrt{\lambda +1} (\lambda +3) f(\lambda)}\\
   &+\frac{2 \log \left(1-r_0^2\right)}{\lambda +3}-\log (r_0)\,,
\end{split}
\end{equation}
where
$
f(\lambda)= \sqrt{2 \lambda  V_p+\lambda +6 V+1}.$

As the RHS of \refeq{eq: impcurve} is a constant, there will exist a conserved quantity $S$,
\begin{equation}
    S=|\sin{\theta_0}|e^
    {-G(r_0)}.
    \label{eq: consQ}
\end{equation}
\\

\begin{figure}
\centering
\includegraphics[width =0.75 \textwidth]{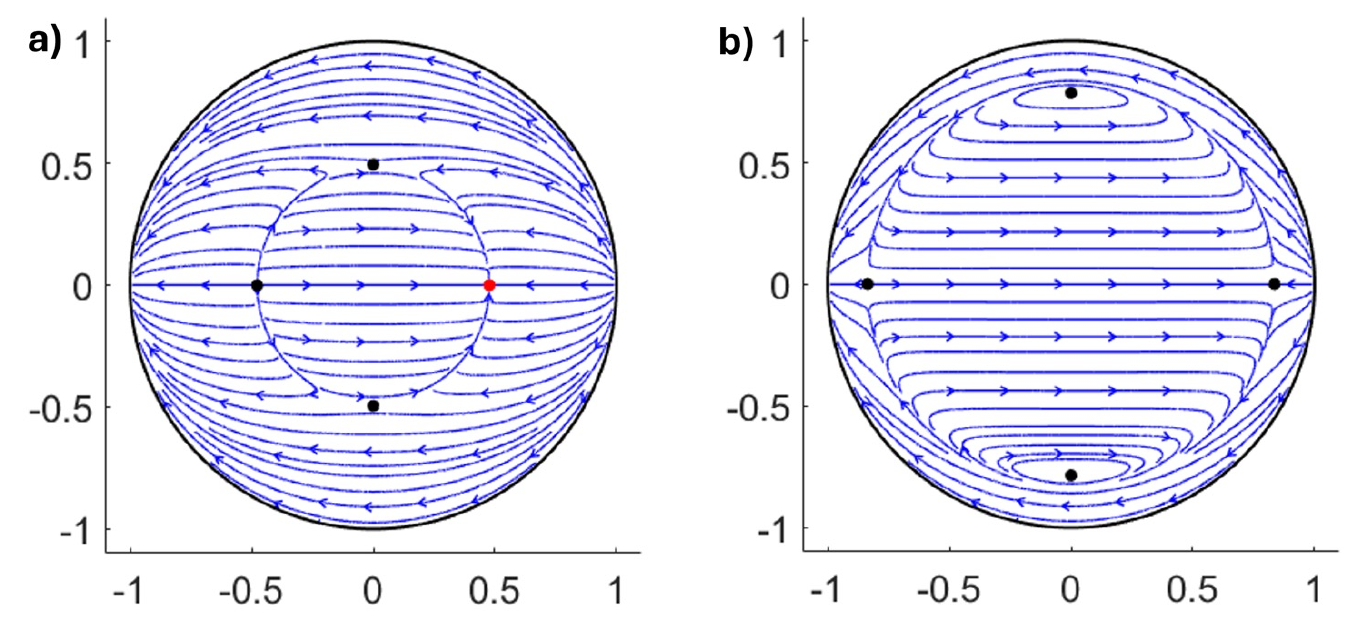}
\caption{{\bf{Phase portrait of the dynamical system \refeq{eq: polarred}}}.  Plot (a) corresponds to $V_p=3$ and $\lambda=2$, therefore having a stable fixed point  at $(r_c,\theta_c)=(0.48,0)$ (red point) and saddles at 
$(r_c,\theta_c)=(0.48,\pi)$ and $(r_p,\theta_p)=(0.49,\pm\frac{\pi}{2})$. Plot (b) corresponds to $V_p=30$ and $\lambda=2$, therefore having centers at $(r_p,\theta_p)=(0.79,\pm\frac{\pi}{2})$ and saddles at  $(r_c,\theta_c)=(0.84,0)$ and $(r_c,\theta_c)(0.84,\pi)$. The system is plot in polar coordinates as depicted in \reffig{fig: stability}. }
\label{fig: flowDip}
\end{figure}

In \reffig{fig: flowDip}, the phase portraits of the dynamical system described by \refeq{eq: polarred} are shown for two regimes.
\reffig{fig: flowDip}(a) corresponds to the regime where $V_p<\frac{(1+\lambda)^2}{2}$, in which there is a stable fixed point at $(r_0,\theta_0)=(r_c,0)$ and an unstable fixed point at $(r_0,\theta_0)=(r_c,\pi)$.
For the region $r_0<r_c$, trajectories approach the radius $r_c$ and then evolve angularly toward the stable fixed point.
There are two saddle points located at $(r_0,\theta_0)=(r_p,\pm \pi/2)$ with $r_p>r_c$.
\reffig{fig: flowDip}(b) corresponds to the regime where $V_p>\frac{(1+\lambda)^2}{2}$.
In this case, there are two centers at $(r_0,\theta_0)=(r_p,\pm \pi/2)$ with $r_p<r_c$.
There are closed trajectories in the region $r_0<r_c$ that encircle these centers.
The fixed points at $(r_0,\theta_0)=(r_c,0)$ and $(r_c,\pi)$ become saddles in this regime.

\subsubsection{Effect of transient domain deformation on the particle motion}
\label{sec: dipoledef}

\begin{figure}
\centering
\includegraphics[width = \textwidth]{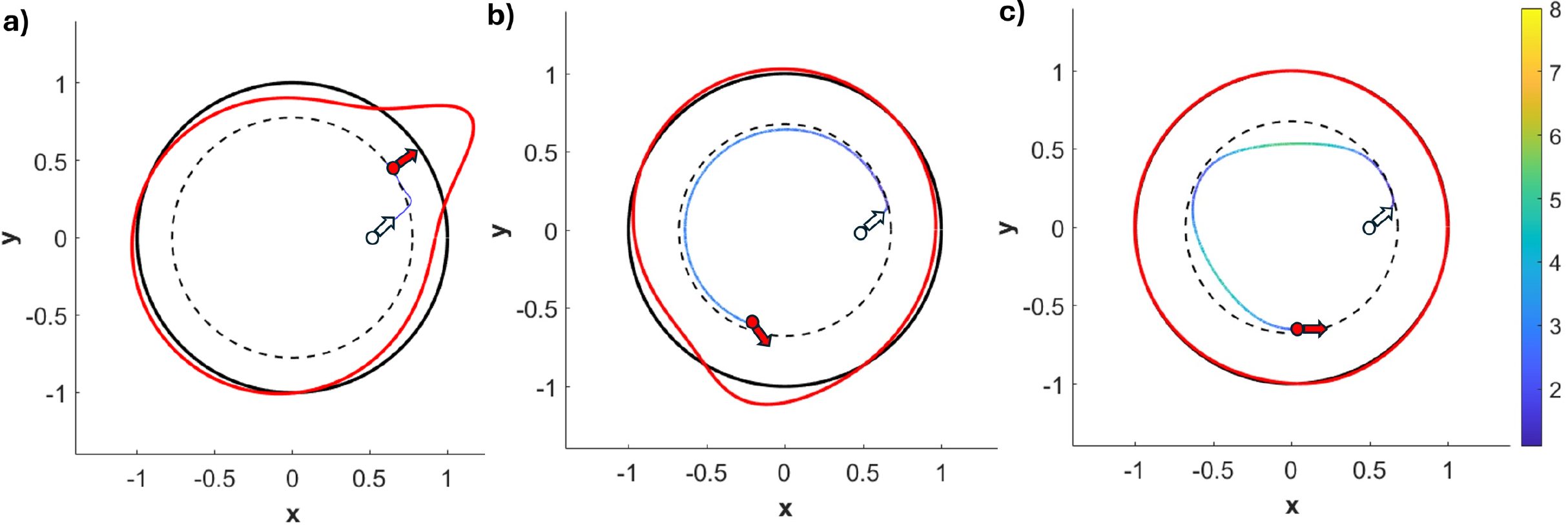}
\caption{ \textbf{Dipole motion in a deformable domain:}
 The initial conditions are $(r_0,\phi_0,\alpha) = (0.5,0,\frac{\pi}{5})$ and $\lambda =1$, with (a) $Ca= 0.015$ and $V_p=20$ (b)  $Ca= 0.01$ and $V_p=10$ and (c)   $Ca= 0.001$ and $V_p=10$. The red curve represents the final shape of the domain. The color bar represents the velocity of the singularity. The white arrow represents the starting point and the red arrow the end point. The black dashed line represents the radial fixed point of \refeq{eq: rc}. The black line represents an undeformed domain.  The simulations are run for one unit of dimensionless time. }
\label{fig numericsource}
\end{figure}

The particle behavior is studied for non-steady shapes by computing the shape evolution in \refeq{eq: shapeevolution} and the dipole migration in \refeq{eq: particleev1} using the second order Adams-Bashforth method. In \reffig{fig numericsource}, a source dipole with self-propulsion velocity $V_p$  
moves inside a deformable domain characterized by the capillary number $Ca$.
In \reffig{fig numericsource}(a), it can be seen that as $\frac{1}{Ca}$ and $V_p$ become comparable in magnitude, the domain relaxes to a steady shape and the dipole approaches the fixed radius given by \refeq{eq: rc} with its orientation pointing radially outwards.  
In \reffig{fig numericsource}(b) and (c), the regime $\frac{1}{Ca} \gg V_p$ is considered.  
In \reffig{fig numericsource}(b), the dipole navigates in nearly circular orbits at a radius close to $r_c$.  
In \reffig{fig numericsource}(c), the capillary number decreases and the behavior approaches the qualitative dynamics of a dipole inside a non-deformable domain. The dipole exhibits increased reorientation as it approaches $r_c$, whereas for $r<r_c$ it swims along nearly straight trajectories.

\subsection{Domain's dynamics  driven by a diffusing stresslet}

Here, we consider an illustrative example of a non-self-propelled force dipole whose orientation undergoes rotational diffusion. This scenario is particularly relevant in biological membranes, where molecular machines are commonly modeled as fluctuating force dipoles \citep{Manneville:2001,mikhailov2015hydrodynamic}. 
In the case $\lambda=1$, the equations of motion  \refeq{eq: stresseqs} with stochastic reorientation \citep{callegari2019numerical,basu2018active} are
\begin{equation}
    \begin{split}
&\frac{d r_0}{dt} =\frac{2 r_0\cos (2(\alpha-\phi_0))}{
\left(r_0^2-1\right)},\\
& \frac{d \phi_0}{dt} = \frac{\left(2- r_0^2\right) \sin (2 (\alpha-\phi_0))}{
\left(r_0^2-1\right)},\\
& \frac{d \alpha}{dt} = -\frac{r^2_{0}(-3+r_0^2)\sin(2(\alpha-\phi_0))}{(r_0^2-1)^2}+ \sqrt{2D_a}\Gamma_\alpha(t).\\
    \end{split} 
    \label{eq: stressrotdif}
\end{equation}
The Gaussian white noise term $\Gamma_\alpha$ has zero mean and correlation $\langle \Gamma_\alpha(t)\Gamma_\alpha(t')\rangle=\delta(t-t')$, with rotational diffusion characterized by the coefficient $D_\alpha$. In the simulations, a short-range radial repulsion with the form $\substack{\frac{\beta}{(1-r_0)^6}}$ is added in the radial equation to prevent the stresslet from leaving the domain.

\begin{figure}
    \centering
    \includegraphics[width = 0.5\textwidth]{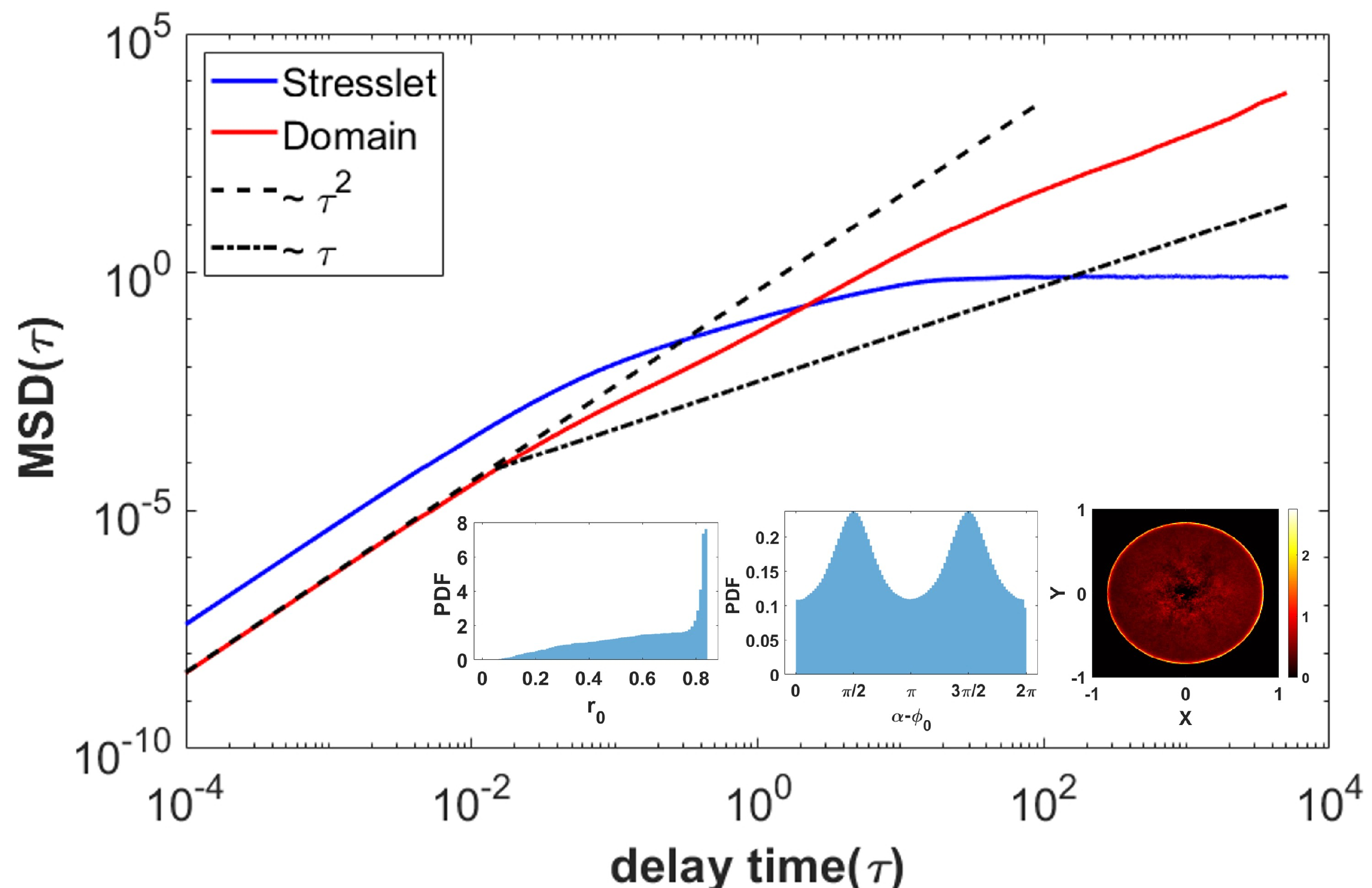}
    \caption{\textbf{Mean square displacement (MSD) of a stresslet inside a domain and the MSD of the domain.} 
     Blue: MSD of a stresslet inside a circular domain. Red: MSD of the domain. Insets (left to right): radial PDF, PDF of the modified angle $\alpha-\phi_0$, and 2D PDF of the stresslet. The modified angle PDF shows a preferred orientation perpendicular to the position vector. The simulation is run for $10^4$ units of dimensionless time     with step size $\Delta t = 10^{-4}$, diffusion constant $D_\alpha = 10$, confinement potential strength $\beta = 10^{-4}$, and viscosity ratio $\lambda = 1$.}
    \label{fig: RWstress}
\end{figure}

In an unbounded fluid, the force-free and torque-free stresslet does not exhibit net translation. Under confinement, however, the stresslet-generated flow interacts with the domain boundary, producing a self-induced advection that drives stresslet motion. When the stresslet orientation $\alpha$ is subject to rotational diffusion, stochastic reorientation competes with this hydrodynamic drift, preventing the stresslet from simply migrating to the boundary. As a result, the stresslet persistently samples the interior of the domain, giving rise to an effective diffusive dynamics.
\reffig{fig: RWstress} shows that the stresslet's mean square displacement (MSD) displays a ballistic regime for $\tau <1/D_\alpha=0.1$ and a diffusive regime for $0.1 < \tau < 10$, where $\tau$ denotes the lag time used to compute the MSD between two points along the trajectory. For $\tau > 10$, the MSD saturates due to confinement by the domain boundary. 

While the stresslet navigates the domain, the instantaneous domain velocity is determined by \refeq{eq: domainvelstress} (assuming a steady-state shape, i.e., shape relaxing fast on the domain displacement time scale). Since the domain velocity depends on the stresslet orientation, the domain also experiences the effects of rotational diffusion. However, in addition to a ballistic (for $\tau < 0.1$) and  a diffusive (for $\tau > 10$) regimes, the domain also exhibits superdiffusion for $0.1 < \tau < 10$.

 The superdiffusive behavior is attributed to strong reorientations of the stresslet within the bulk region. In the inset of \reffig{fig: RWstress}, the probability density function of the relative angle $\alpha - \phi_0$ reveals the stresslet's tendency to orient perpendicular to its radial position vector. This preferred orientation induces a positive bias in the radial velocity term of \refeq{eq: domainvelstress}. For $\tau > 10$, the domain motion becomes diffusive since the stresslet is trapped near the boundary and its dynamics are dominated by the radial repulsion. In the case of a passive source dipole, the domain's MSD also presents similar behavior.

\label{sec: RW}

\section{Conclusions and outlook}
\label{sec: Conclusion}
This paper presents a theoretical analysis of two-dimensional flow driven by an active particle confined within a deformable, initially circular droplet. The active particle is modeled as a Stokes-flow singularity. A perturbation analysis is developed in the limit of small capillary number, where active stresses are weak relative to line tension and the domain shape remains close to circular. While this study is motivated by biological membranes, where membrane-bound biomolecular machines may generate active flows that drive the motion and active diffusion of lipid domains, it also contributes more broadly to the theory of two-dimensional Stokes flows generated by hydrodynamic singularities in confined geometries, e.g., channels \citep{crowdy2013stokes}, wedge-shaped domains \citep{crowdy2017analytical}, rigid circular boundaries \citep{avudainayagam1987no,palaniappan2001interior}, annular domains \citep{hackborn2000analysis}, and deformable planar interfaces \citep{crowdy2011two}. Here, we address the previously unexplored problem of an active singularity enclosed within a deformable circular domain, where the flow, particle motion, and domain shape evolve in a coupled manner.

Explicit results are obtained for a source dipole and a stresslet. We find that these force-free singularities induce both domain deformation and net translational motion. For a source dipole, the domain translates along the dipole axis with a velocity that is independent of the particle's position within the domain. In contrast, a stresslet does not propel the domain along its own axis of symmetry. Rather, the induced domain velocity depends sensitively on the stresslet position and orientation. These behaviors are qualitatively similar to those reported for an active particle confined within three-dimensional droplets \citep{kawakami2025migration} and Hele--Shaw droplets \citep{kawakami2025microswimmer}.

For a motile source dipole, confinement can induce periodic particle motions. Although a source dipole generates an irrotational flow in unbounded space, confinement introduces vorticity, which drives particle reorientation and can lead to periodic orbiting within the domain. Depending on the propulsion speed 
$V_p$ and viscosity ratio $\lambda$, the source dipole may exhibit either a stable fixed point or persistent orbital motion. In contrast, the confinement-induced flow generated by a stresslet does not support stable fixed points or closed orbits; its dynamics are simpler, with the particle monotonically swimming outward toward the boundary.
When the stresslet undergoes rotational diffusion within a domain, its deterministic reorientation and tendency to align perpendicular to its position vector lead to transient superdiffusion of the domain. This superdiffusive regime is also present for a stresslet in a three-dimensional drop \citep{kawakami2025migration}. 

This work examines two-dimensional flows generated by an active particle confined to an incompressible-fluid interface, such as a lipid membrane or a surfactant-laden interface \citep{manikantan2020surfactant}. Coupling to the surrounding three-dimensional fluid becomes important when the Saffman--Delbrück length is comparable to or smaller than the characteristic size of the system, in which case viscous dissipation in the bulk fluid significantly influences the interfacial dynamics.{ Such coupling is also relevant for particles that protrude into the three-dimensional fluids adjacent to the membrane \citep{danov2000viscous,fischer2006viscous, stone2015mobility}.} Nevertheless, the present analysis provides a fundamental framework for addressing these more general problems and serves as a useful starting point for future studies of active particles at fluid interfaces \citep{deng2022active,deng2023interfacial}.

The present analysis assumes a point-like active particle confined to a planar membrane. Future work could relax both assumptions by considering finite-size particles and curved membrane geometries. The mobility of rods \citep{levine2004mobility} and arbitrary-shaped inclusions \citep{camley2013diffusion} has previously been investigated in planar membranes, while particle dynamics on curved membranes have also been analyzed \citep{Henle:2010,Arroyo:2013,Sigurdsson:2016,shi2022hydrodynamics,vona2025rotational}. These extensions would broaden the applicability of the present theory to more general biological membrane systems. Moreover, the present study assumes that the membrane is a two-dimensional incompressible interface. The more general case of a fluid interface that can change area has also been extensively studied \citep{dorr2016drag, pourali2021drag,zhou2022drag,zhou2025drag, maldarelli2022continuum}, particularly in the context of particle-stabilized emulsions and colloidal self-assembly \citep{McGorty2010Colloidal, liu2018capillary}. Future work could consider active particles trapped at interfaces with complex rheology \citep{vermant:2012}.

The developed theory could be used to assess the activity of molecular machines by analyzing the diffusion of domains in biomimetic membranes that can be readily observed with optical microscopy. This, in turn, may provide new insights into non-equilibrium effects on membrane dynamics. Beyond non-motile biomolecular machines, advances in nanoengineering \citep{Balzani:2000,Kay:2015,feng2021molecular, zhou:2021} may make self-propelled nanomotors and active transport on cellular scales a reality.

\section*{Declaration of Interests}
The authors report no conflict of interest.

\section*{Acknowledgments}
This research  was supported by NSF award DMR 2415770.

\appendix
\section{Fundamental solutions of the Stokes equation in 2D}
\label{appendix: 2D singularity solutions}
\noindent A complex basis for the Stokes equation 
\begin{align}
    -\bnab  p^{\pm}_{m,\s}+\nabla^2 \bv^{\pm}_{m,\s} = 0, 
\end{align}
 is given by velocities, with $m \in \mathbb{Z}$,
\begin{flalign}
	\bu_{m,0}^- &= \begin{cases}
		-r^{-1} \phihat &m=0\\
		 r^{-|m|-1}e^{i m \phi} (im\rhat+|m|\phihat)&|m|\geq 1
	\end{cases}\label{Eq:ComplexBasis1}&&\\
	\bu_{m,1}^- &= \begin{cases}
		0 &m=0,1\\
		r^{-|m|+1}e^{i m \phi} (im\rhat+(|m|-2)\phihat)&|m|\geq 2
	\end{cases}\label{Eq:ComplexBasis2}&&\\
	\bu_{m,0}^+ &= \begin{cases}
		0 &m=0\\
		r^{|m|-1}e^{i m \phi} (im\rhat-|m|\phihat)&|m|\geq 1
	\end{cases}\label{Eq:ComplexBasis3}&&\\
	\bu_{m,1}^+ &= \begin{cases}
		-2r \phihat &m=0\\
		r^{|m|+1}e^{i m \phi} (im\rhat-(|m|+2)\phihat)&|m|\geq 1
	\end{cases}\label{Eq:ComplexBasis4}
\end{flalign}
and corresponding pressures
\begin{flalign}
&p_{m,0}^\pm = 0,  &&\\
&p_{m,1}^\pm = 4 i (|m|\pm 1)\sgn(m) r^{\pm|m|}e^{i m \phi}.
\end{flalign}

\noindent  All the flow velocities are incompressible,
\begin{equation}
\bnab \cdot\bv^{\pm}_{m,\s} = 0.
\end{equation}
The superscript $\pm$ denotes solutions that are regular either at infinity $(-)$ or at the origin $(+)$, and the index $\s$ distinguishes between irrotational $(\s=0)$ and rotational $(\s=1)$ modes.

\noindent The flows will have associated angular velocity 

\begin{equation}
   \begin{split}
\bm{\omega}^+_{m,0} = \frac{1}{2}\nabla \times \bu_{m,0}^+ &= \begin{cases}
		0 &m=0\\
		0&|m|\geq 1
	\end{cases}\\
	\bm{\omega}^+_{m,1} = \frac{1}{2} \nabla \times \bu_{m,1}^+ &= \begin{cases}
		-2 \zhat  &m=0\\
		-2r_0^{|m|}e^{i m \phi_0} (|m|+1)\zhat&|m|\geq 1
	\end{cases}
\end{split} 
\label{eq: flowcurl}
\end{equation}

\noindent 

The domain velocity is given by \refeq{eq: domainvelocity}. The surface integral is evaluated to be
\begin{equation}
	\frac{1}{\pi}\int_0^{2\pi} (\bv_{m,\s}^-(1,\phi) \cdot \rhat)\rhat d\phi = \begin{cases}
		0 & |m| \neq 1 \text{ or }\s=1\\
		(i,-1)& m=1,\s=0\\
		(-i,-1)&m=-1,\s=0
	\end{cases}
    \label{eq: intflow}
\end{equation}

\noindent The bulk hydrodynamic stress is given by
\begin{flalign}
	&\bT = -p \bm{\delta}+ \bnab \bu +(\bnab \bu)^T.
\end{flalign}
For each basis element, the bulk hydrodynamic stress is given by
\begin{flalign}
	& \bT_{m,0}^- =  2 e^{i m \phi}r^{-|m|-2}\begin{bmatrix}
		-i m(|m|+1)&-(m^2+|m|)\\-(m^2+|m|)&i m(|m|+1)
	\end{bmatrix} \quad |m|\geq 1,\\
		&\bT_{m,1}^- =  2 e^{i m \phi}r^{-|m|}\begin{bmatrix}-
			i (|m|-1)(m+2\sgn(m))&-(m^2-|m|)\\-(m^2-|m|)&i (|m|-1)(m-2\sgn(m))
		\end{bmatrix}\quad |m|\geq 2,\\
		&\bT_{m,0}^+ =  2 e^{i m \phi}r^{|m|-2}\begin{bmatrix}
			i m(|m|-1)&-(m^2-|m|)\\-(m^2-|m|)&-i m(|m|-1)
		\end{bmatrix} \quad |m|\geq 1,\\
		&\bT_{m,1}^+ =  2 e^{i m \phi}r^{|m|}\begin{bmatrix}
			i (|m|+1)(m-2\sgn(m))&-(m^2+|m|)\\-(m^2+|m|)&-i (|m|+1)(m+2\sgn(m))
		\end{bmatrix} \quad |m|\geq 1,
\end{flalign}
with special cases
\begin{flalign}
  &{\bT}_{0,0}^- =  \frac{2}{r^2}\begin{bmatrix}
		0&1\\1&0
	\end{bmatrix},\\
&{\bT}_{0,1}^+ =  \begin{bmatrix}
		0&0\\0&0
	\end{bmatrix}.
\end{flalign}
 The normal stress on a circular domain, evaluated at $r=1$, is given by
\begin{flalign}
	&\rhat \cdot \bT_{m,0}^-|_{r=1} = 2 e^{i m \phi}\left[-im(|m|+1)\rhat-(m^2+|m|)\phihat\right] \quad |m|\geq 1,\\
	&\rhat \cdot \bT_{m,1}^-|_{r=1} =  2 e^{i m \phi}\left[-i(|m|-1)(m+2\sgn(m))\rhat-(m^2-|m|)\phihat\right] \quad |m|\geq 2,\\
	&\rhat \cdot \bT_{m,0}^+|_{r=1} = 2 e^{i m \phi}\left[im(|m|-1)\rhat-(m^2-|m|)\phihat\right] \quad |m| \geq 1, \\
	&\rhat \cdot \bT_{m,1}^+|_{r=1} =  2 e^{i m \phi}\left[i(|m|+1)(m-2\sgn(m))\rhat-(m^2+|m|)\phihat\right] \quad |m|\geq 1,\\
\end{flalign}
with special cases
\begin{flalign}
    &\rhat \cdot \bT_{0,0}^-|_{r=1} = 2\phihat,\\
    &\rhat \cdot \bT_{0,1}^+|_{r=1} = \textbf{0}.
\end{flalign}
Given the flow
\begin{flalign}
	\bu = \sum_{m=-\infty}^\infty \sum_{\s=0,1}c_{m,\s}^\pm\bu^\pm_{m,\s},
\end{flalign}
the normal stress on a circular interface at $r=1$ is given by
\begin{flalign}
	& \nhat \cdot \bT  = \sum_{m=-\infty}^\infty \textbf{A}^\pm_m \begin{bmatrix}
		c_{m,0}^\pm\\c_{m,1}^\pm
	\end{bmatrix} e^{i m \phi},
\end{flalign}
with 
\begin{flalign}
	& \textbf{A}^-_m = \begin{bmatrix}
		-2i m (|m|+1) \rhat&-2i(|m|-1)(m+2\sgn(m))\rhat\\
		-2(m^2+|m|)\phihat & -2(m^2-|m|) \phihat
	\end{bmatrix},\\
	&\textbf{A}^+_m = \begin{bmatrix}
		2im(|m|-1)\rhat&2i(|m|+1)(m-2\sgn(m))\rhat\\
		-2(m^2-|m|)\phihat & -2(m^2+|m|)\phihat
	\end{bmatrix},
\end{flalign}
 for $|m|\geq 1$ and special cases

\begin{flalign}
	& \textbf{A}^-_0  = \begin{bmatrix}
	 \textbf{0}& \textbf{0}\\
		2\phihat &  \textbf{0}
	\end{bmatrix},\\
	&\textbf{A}^+_0 = \begin{bmatrix}
		 \textbf{0} &  \textbf{0}\\
		 \textbf{0} &  \textbf{0}
	\end{bmatrix}.
\end{flalign}

\section{Transform}
\label{appendix: FJ transform}

\noindent Consider a singularity flow given by $\bv = c_{m,\s}^{-} \bv^{-}_{m,\s}+c_{-m,\s}^{-} \bv^{-}_{-m,\s}$ in its own coordinate system with $m \in \mathbb{Z}$. The singularity located at $\bxz=(x_0,\phi_0)$ is transformed to the basis $m',\s'$ centered at the origin given coefficients $ c_{m',\s'}^{\displaced,-}$ for regions $x>x_0$ and $ c_{m',\s'}^{\displaced,+}$ for regions $x<x_0$ in the form
\begin{equation}
\bv^{\displaced}(\bx,\bx_0) =\summsp  c_{m',\s'}^{\displaced,+}(\bxz)\bu_{m',\s'}^+(\bx)H(|\bx_0|-|\bx|)+c_{m',\s'}^{\displaced,-}(\bxz)\bu_{m',\s'}^-(\bx)H(|\bx|-|\bx_0|),
\end{equation}
with the transformed coefficients given by
\begin{flalign}
    &c_{m',\s'}^{\displaced,+}=c_{-m,\s}^{-}c_{-m,\s,m',\s'}^{-+}+c_{m,\s}^{-}c_{m,\s,m',\s'}^{-+},\\ &c_{m',\s'}^{\displaced,-}= c_{-m,\s}^{-}c_{-m,\s,m',\s'}^{--}+c_{m,\s}^{-}c_{m,\s,m',\s'}^{--}.
\end{flalign}

The transform coefficients for the outer region  are
\begin{flalign}
c_{0,0,m',0}^{--}&=\begin{cases}
		1&m'=0\\
		-\frac{r_0^{|m'|}}{2|m'|}e^{-im'\phi_0}&|m'|>0
	\end{cases}&&\\
	c_{m,0,m',0}^{--}&=\begin{cases}
        0&|m'|<|m|,|m|>0\\
		1&|m|=|m'|,| m|>0\\
		\binom{|m'|-1}{|m|-1} r_0^{|m'|-|m|}\delta_{\sgn(m),\sgn(m')}e^{-i\sgn(m')(|m'|-|m|)\phi_0}&|m'|>|m|,|m|>0
	\end{cases}  &&\\
    c_{m,0,m',1}^{--}&=0&&\\
	c_{m,1,m',0}^{--}&=\begin{cases}
		0&|m'|<|m|-1\\
		-r_0\delta_{\sgn(m),\sgn(m')}e^{i \sgn(m') \phi_0} & |m'|=|m|-1\\
		-\binom{|m'|-1}{|m|-2}r_0^{|m'|-|m|+2} \delta_{\sgn(m),\sgn(m')}e^{-i\sgn(m')(|m'|-|m|)\phi_0}&|m'|\geq |m|
	\end{cases} &&\\
		c_{m,1,m',1}^{--}&=\begin{cases}
		0&|m'|<|m|\\
		1&|m|=|m'|\\
		r_0^{|m'|-|m|}\binom{|m'|-2}{|m|-2}\delta_{\sgn(m),\sgn(m')} e^{-i\sgn(m')(|m'|-|m|)\phi_0}&|m'|\geq |m|+1
	\end{cases}&&
    \end{flalign}
and the transform coefficients for the inner region are
\begin{flalign}
	c_{0,0,m',0}^{-+}&=\begin{cases}
		0&m'=0\\
		-\frac{r_0^{-|m'|}}{2|m'|}e^{-im'\phi_0}&|m'|>0
	\end{cases}&&\\
	c_{m,0,m',0}^{-+}&=\begin{cases}
		0& \substack{|m'|=0,| m|>0}\\
		\substack{(-1)^{|m|}\binom{|m'|+|m|-1}{|m|-1} r_0^{-(|m'|+|m|)}\delta_{\sgn(m),-\sgn(m')}e^{-i\sgn(m')(|m'|+|m|)\phi_0}}&\substack{|m'|\geq 1,|m|>0}
	\end{cases}&&\\
    c_{m,0,m',1}^{-+}&=0&&\\
	c_{m,1,m',0}^{-+}&=\begin{cases}
    \begin{aligned}
    &\substack{(-1)^{|m|}r_0^{-|m|+1}(|m|-1)\delta_{\sgn(m),-\sgn(m')}e^{-i\sgn(m')(1+|m|)\phi_0}+}\\
    & \substack{-(-1)^{|m|}r_0^{-|m|+1}\delta_{\sgn(m),\sgn(m')}e^{i\sgn(m')(|m|-1)\phi_0}} \\
    \end{aligned}
       & \substack{|m'|=1, |m|\geq 2}\\
		\substack{(-1)^{|m|}\binom{|m|+|m'|-2}{|m|-2}r_0^{-|m'|-|m|+2} \delta_{\sgn(m),-\sgn(m')}e^{-i\sgn(m')(|m'|+|m|)\phi_0}}&
        \substack{|m'|>1, |m|\geq 2}\\
	\end{cases}&&\\
c_{m,1,m',1}^{-+}&=\begin{cases}
        \substack{-(-1)^{|m|}r_0^{-|m|}\binom{|m|-1}{|m|-2} e^{-i\sgn(m)|m|\phi_0}}& \substack{|m'| = 0,|m|\geq2}\\
		\substack{-(-1)^{|m|}r_0^{-(|m'|+|m|)}\binom{|m|+|m'|-1}{|m|-2}\delta_{\sgn(m),-\sgn(m')} e^{-i\sgn(m')(|m'|+|m|)\phi_0}}& \substack{|m'|> 0,|m|\geq2}
	\end{cases}&&
\end{flalign}
\section{Coefficients for the induced flows of different singularties}

\subsection{ Stresslet}
A stresslet flow is described by \citep{lushi2014fluid} $\textbf{f}(\bx) =\bnab \cdot\left( 4\pi( \phat\phat) \delta(\bx - \bf{x_0})\right)$ and $g(\bx)=0$ in \refeq{eq: StokesEqinside}, with $\phat$ oriented at an angle $\alpha$ relative to the x-axis. The expression for the unbounded  stresslet flow in a coordinate system centered at the particle is given by
\begin{equation}
    \bv^{\undisplaced}(\bx) = \frac{1}{\lambda r}(2(\phat\cdot \rhat)^2\rhat-\rhat)=\frac{1}{\lambda}\frac{\cos(2(\phi-\alpha))}{r}\rhat.
    \label{eq: closestress}
\end{equation}
This flow  is illustrated in \reffig{fig: stressletflow}(a). The flow coefficients
in the velocity representation \refeq{Eq: undisplacedsingularity} are 
\begin{equation}
     c_{-2,1}^{\undisplaced}=\frac{1}{2\lambda}e^{i(2\alpha+\frac{\pi}{2})},\quad  c_{2,1}^{\undisplaced}=\frac{1}{2\lambda}e^{-i(2\alpha+\frac{\pi}{2})}.
\end{equation}
Using the transform in \refapp{appendix: FJ transform}, the coefficients in \refeq{eq: displacedsingularity} for the stresslet-generated flow in the coordinate system centered at the domain are given by 
\begin{equation}
    \begin{split}
  &c_{0,0}^{\displaced,-,\stresslet} =0, \quad
  c_{m,0}^{\displaced,-,\stresslet} =-\frac{1}{2\lambda}r_0^{|m|}K \quad |m| \geq 1, \\
    &c_{0,1}^{\displaced,-,\stresslet} =c_{1,1}^{\displaced,-,\stresslet}=0, \quad c_{m,1}^{\displaced,-,\stresslet} =\frac{1}{2\lambda}r_0^{|m|-2}K \quad  |m|\geq 2,\\
     &K \equiv e^{-i \sgn(m) \left(2\alpha+(| m|-2)\phi_0 +\frac{\pi }{2}\right)},\\
      &c_{0,0}^{\displaced,+,\stresslet} =0, \quad \quad   c_{1,0}^{\displaced,+,\stresslet} =\frac{1}{2\lambda r_0}Y-\frac{1}{2\lambda r_0}K \quad |m|= 1,\\
 & c_{m,0}^{\displaced,+,\stresslet} =\frac{1}{2\lambda}r_0^{-|m|}Y \quad |m| \geq 2, \\
    &c_{0,1}^{\displaced,+,\stresslet} =-\frac{r_0^{-2}}{\lambda}\sin(2(\phi_0-\alpha)), \quad c_{m,1}^{\displaced,+,\stresslet} =-\frac{1}{2\lambda}r_0^{-|m|-2}Y\quad  |m|\geq 1,\\
     &Y \equiv e^{-i \sgn(m) \left((| m|+2)\phi_0-2\alpha -\frac{\pi }{2}\right)}.\\
    \end{split}
    \label{eq: coffstress}
\end{equation}

 The coefficients for the confinement-induced flow are obtained by substituting \refeq{eq: coffstress} into \refeq{eq: coffout} and \refeq{eq: coffin}
\begin{equation}
\begin{split}
      &c_{m,0}^{\out,\stresslet} =-\frac{4 r_0^{|m|}K+if_m(-1+|m|)\sgn(m)}{4(1+\lambda)}\quad|m|> 1,\\
      &c_{m,1}^{\out,\stresslet} =\frac{4 r_0^{|m|-2}K+if_m(1+|m|)\sgn(m)}{4(1+\lambda)}\quad|m| > 1,\\
        &c_{\pm 1,0}^{\out,\stresslet} =-\frac{r_0e^{\mp i((2\alpha+\frac{\pi}{2})-\phi_0)}}{(1+\lambda)}, \quad c_{ 0,0}^{\out,\stresslet} =0 ,\\
      &c_{m,0}^{\ins,\stresslet} =\frac{-2K(\lambda -1) r_0^{| m|-2 } \left(\left(r_0^2-1\right) | m| +r_0^2\right)+i f_m \lambda \sgn(m)(1+|m|) }{4\lambda (\lambda +1)}\quad|m|>1,\\
       &c_{m,0}^{\ins,\stresslet} =\frac{2K(\lambda -1) r_0^{| m|-2 } \left(\left(r_0^2-1\right) | m| +1\right)-i f_m \lambda \sgn(m)(-1+|m|) }{4\lambda (\lambda +1)}\quad|m|>1,\\
         &c_{\pm 1,0}^{\ins,\stresslet} =-e^{\mp i(2\alpha+\frac{\pi}{2}-\phi_0)}\frac{-1+\lambda}{(1+\lambda)\lambda}r_0, \quad
         c_{\pm 1,1}^{\ins,\stresslet} =e^{\mp i(2\alpha+\frac{\pi}{2}-\phi_0)}\frac{-1+\lambda}{2(1+\lambda)\lambda}r_0, \quad  c_{0 ,1}^{\ins,\stresslet} =0.\\
\end{split}
\label{eq: coffstressimag}
\end{equation}

Setting the normal velocity to zero, $c_{m,0}^{\out,\stresslet}+c_{m,1}^{\out,\stresslet}=0$,  in \refeq{eq: coffstressimag}, and solving for the mode coefficients for $|m| \geq 2$, yields for the steady domain shape 
\begin{equation}
\begin{split}
    & f_m = -2 i \left(r_0^2-1\right) \sgn(m) r_0^{| m| -2}K,\\
     &c_{m,0}^{\ins,\stresslet,\steadyshape} =\frac{r_0^{| m| -2} \left(\left(r_0^2-\lambda \right)+|m|\left(r_0^2-1\right)\right) K}{2\lambda (\lambda +1)}, \quad 
       c_{m,1}^{\ins,\stresslet,\steadyshape} =\frac{r_0^{| m| -2} \left((1-r_0^2)| m| +\lambda  r_0^2-1\right)K}{2\lambda(\lambda +1)},\\
      &c_{m,0}^{\out,\stresslet,\steadyshape} =-\frac{  r_0^{| m| -2} \left(\left(r_0^2-1\right)| m| +1+ r_0^2\right)K}{2(\lambda +1)}, \quad c_{m,1}^{\out,\stresslet,\steadyshape} =\frac{ r_0^{| m| -2} \left(\left(r_0^2-1\right)| m| +1+ r_0^2\right) K}{2(\lambda +1)},
\end{split}
\label{eq: coffstresseq}
\end{equation}
with the same coefficients for  $|m| =1$ as those in \refeq{eq: coffstressimag}.
\subsection{Source dipole}
A source dipole flow corresponds to $\textbf{f}(\bx)=0$ and $g(\bx) = - 4\pi\phat \cdot \bnab \delta(\bx - \bf{x_0})$ in \refeq{eq: StokesEqinside} where $\phat=[\cos{(\alpha)},\sin{(\alpha)}]^{T}$. The expression for the unbounded source dipole flow in a coordinate system centered at the particle is given by
 \begin{equation}
    \bv^{\undisplaced}(r,\phi) =\frac{2}{r^2}\left(\cos(\phi-\alpha)\rhat+\sin(\phi-\alpha)\phihat\right).
    \label{eq: dipflow}
\end{equation}
The flow coefficients in the velocity representation \refeq{Eq: undisplacedsingularity} are 
\begin{equation}
  c_{-1,0}^{\undisplaced}=e^{i(\alpha+\frac{\pi}{2})},\quad   c_{1,0}^{\undisplaced}=e^{-i(\alpha+\frac{\pi}{2})}.
\end{equation}
Using the transform in \refapp{appendix: FJ transform}, the coefficients in \refeq{eq: displacedsingularity} for the dipole-generated flow  in the coordinate system centered at the domain are given by 
\begin{equation}
\begin{split}
& c_{0,0}^{\displaced,-,\dipole} =0\,,\quad
  c_{m,0}^{\displaced,-,\dipole} =r_0^{|m|-1}Q \quad|m|\geq1, \quad Q\equiv e^{-i\sgn(m)\left((\alpha+\frac{\pi}{2})+(|m|-1)\phi_0\right)},\\
&c_{0,0}^{\displaced,+,\dipole} =0\,,\quad
  c_{m,0}^{\displaced,+,\dipole} =-r_0^{-|m|-1}W \quad|m|\geq1, \quad W\equiv e^{i\sgn(m)\left((\alpha+\frac{\pi}{2})+(|m|+1)\phi_0\right)},\\
  &c_{m,1}^{\displaced,\pm,\dipole} =0.
  \end{split}
  \label{eq: coffdip}
\end{equation}
The coefficients for the confinement-induced flow are obtained by substituting \refeq{eq: coffdip} into \refeq{eq: coffout} and \refeq{eq: coffin}
\begin{equation}
\begin{split}
      &c_{m,0}^{\out,\dipole} =\frac{8\lambda r_0^{|m|-1}Q+if_m(1-|m|)\sgn(m)}{4(1+\lambda)},\quad
      c_{m,1}^{\out,\dipole} =\frac{if_m(1+|m|)\sgn(m)}{4(1+\lambda)}\quad|m|>1,\\
       &c_{\pm1,0}^{\out,\dipole} =\frac{2\lambda}{1+\lambda}e^{\mp i(\alpha+\frac{\pi}{2})}, \quad c_{0,0}^{\out,\dipole} =0, \\
      &c_{m,0}^{\ins,\dipole} =(1+|m|)\frac{4r_0^{|m|-1}Q(-1+\lambda)+if_m\sgn(m)}{4(1+\lambda)}\quad|m|>1,\\
       &c_{m,1}^{\ins,\dipole} =\frac{4(1-\lambda) | m|Q  r_0^{| m|
   -1}- if_m (|m|-1)\sgn(m)}{4 (\lambda +1)}\quad|m|>1,\\
        & c_{\pm1,0}^{\ins,\dipole} =2e^{\mp i(\alpha+\frac{\pi}{2})}\frac{\lambda-1}{1+\lambda}, \quad c_{\pm1,1}^{\ins,\dipole} =e^{\mp i(\alpha+\frac{\pi}{2})}\frac{1-\lambda}{1+\lambda},\quad  c_{0,1}^{\ins,\dipole} =0.
    \end{split}
    \label{eq: coffdipimage}
\end{equation}

Setting  the normal velocity to zero, $c_{m,0}^{\out,\dipole}+c_{m,1}^{\out,\dipole}=0$,  in \refeq{eq: coffdipimage}, and solving for the mode coefficients for $|m| \geq 2$, yields for the steady domain shape 
\begin{equation}
\begin{split}
    &f_m^\steadyshape = 4i\sgn(m)\lambda r_0^{|m|-1}Q, \\
      &c_{m,0}^{\out,\dipole,\steadyshape} =\frac{(1+|m|)\lambda r_0^{|m|-1}Q}{(\lambda +1)},  \quad
      c_{m,1}^{\out,\dipole,\steadyshape} =- \frac{(1+|m|)\lambda r_0^{|m|-1}Q}{(\lambda +1)},\\
      &c_{m,0}^{\ins,\dipole,\steadyshape} =-(1+|m|)\frac{r_0^{|m|-1}Q}{(\lambda +1)},\quad
     c_{m,1}^{\ins,\dipole,\steadyshape} =\frac{r_0^{|m|-1}Q(-\lambda+|m|)}{(1+\lambda)},\\
\end{split}
\label{eq: coffdipeq}
\end{equation}
with the same coefficients for  $|m| =1$ as those in \refeq{eq: coffdipimage}.

\section{Analysis of the dynamical system for the swimming trajectory}
The dynamical equations for a stresslet inside a non-deformable domain are found in \refeq{eq: stresseqs}. The linear stability of this system is done in \refsec{sec: jacobianStress}. \noindent The simplified dynamical equations for a source dipole inside a circular domain are found in \refeq{eq: polarred}. In \refsec{sec: proofF}, the radial equation of \refeq{eq: polarred} is proven to have an unique zero. In \refsec{sec: proofH}, the angular equation of \refeq{eq: polarred} is proven to have an unique zero. Similarly, the linear stability of the system is found in \refsec{sec: jacobian}. 

\label{appendix: proofs}
\subsection{Linear stability analysis of the system in \refeq{eq: stresseqs}}
\label{sec: jacobianStress}
Consider the system
\begin{equation}
    \begin{split}
&F(r_0,\theta_0,\phi_0)\equiv \frac{d r_0}{dt} =V_p\cos(\alpha-\phi_0)+\frac{2 r_0\cos (2(\alpha-\phi_0))}{
\lambda\left(r_0^2-1\right)},\\
&G(r_0,\theta_0,\phi_0)\equiv  \frac{d \phi_0}{dt} = \frac{V_p}{r_0}\sin(\alpha-\phi_0)+\frac{2\left(1+\lambda(1- r_0^2)\right) \sin (2 (\alpha-\phi_0))}{\lambda(\lambda +1)
\left(r_0^2-1\right)},\\
& H(r_0,\theta_0,\phi_0)\equiv \frac{d \alpha}{dt} = -\frac{2r^2_{0}(-1+\lambda(-2+r_0^2))\sin(2(\alpha-\phi_0))}{\lambda(r_0^2-1)^2(1+\lambda))}.
    \end{split} 
\end{equation}
 Fixed points can occur for the angular equations if $\alpha-\phi_0=0$ or $\alpha-\phi_0=\pi$. For the case $\alpha-\phi_0=\pi$, the radial equation cannot be zero as $V_p>0$. Then for $\alpha-\phi_0=0$, the radial critical point is given by the positive root
 \begin{equation}
     r_c=(-1+\sqrt{1+\lambda^2V_p^2})/(\lambda V_p),\quad 0<r_c<1.
 \end{equation}
 Fixed points  occur at  $(r_0,\phi_0,\alpha_0)=(r_c,\phi_c,\phi_c)$. The Jacobian of the system evaluated at $r_0=r_c$ and  $\alpha-\phi_0=0$ is given by

 \begin{equation}
    J(r_c,0,0)=\begin{bmatrix}
        \frac{\partial F}{\partial r_0}(r_c,0,0) &&  0 && 0\\
        0 && \frac{\partial G}{\partial \phi_0}(r_c,0,0) &&  \frac{\partial G}{\partial \alpha}(r_c,0,0)\\
        0 && \frac{\partial H}{\partial \phi_0}(r_c,0,0) &&  \frac{\partial H}{\partial \alpha}(r_c,0,0)\\\\
    \end{bmatrix}.
\end{equation}
Due to the odd symmetry of the derivatives with respect to $\phi_0$ and $\alpha_0$, the Jacobian can be simplified to
 \begin{equation}
    J(r_c,0,0)=\begin{bmatrix}
        \frac{\partial F}{\partial r_0}(r_c,0,0) &&  0 && 0\\
        0 && -\frac{\partial G}{\partial \alpha}(r_c,0,0) &&  \frac{\partial G}{\partial \alpha}(r_c,0,0)\\
        0 && -\frac{\partial H}{\partial \alpha}(r_c,0,0) &&  \frac{\partial H}{\partial \alpha}(r_c,0,0)\\\\
    \end{bmatrix}.
\end{equation}
One eigenvalue is given by 
\begin{equation}
    \Lambda_1=\frac{\partial F}{\partial r_0}(r_c,0,0)=-2\frac{1+r_c^2}{\lambda(r_c^2-1)^2}<0,
\end{equation}
 which is always negative.
Another eigenvalue will be $\Lambda_2=0$, as the lower matrix block has zero determinant. Therefore substracting the known eigenvalues from the trace, the third eigenvalue is given by 

\begin{equation}
   \Lambda_3 = -\frac{\partial G}{\partial \alpha}(r_c,0,0)+\frac{\partial H}{\partial \alpha}(r_c,0,0)=2\frac{1+r_c^2}{\lambda(r_c^2-1)^2}>0.
\end{equation}
 As $\Lambda_1<0$ and $\Lambda_3>0$, the fixed point that are given by $r_0=r_c$ and $\alpha-\phi_0=0$ will be a saddle.  
 
\subsection{Condition for a unique zero of the radial equation in \refeq{eq: polarred}}\label{sec: proofF}
Take the function 
\begin{equation}
F(r_0)=\left(V_p-2\frac{2+(1+\lambda)r_0^2}{(1+\lambda)(r_0^2-1)^2}\right).
\end{equation}
 \noindent \textbf{Proposition}: Assume $\lambda>0$ and the regime $V_p>\frac{4}{1+\lambda}$. Then $F(r_0)$ has an unique zero in the range $0<r_0<1$.
 \newline
\textbf{Proof}
The evaluation of the function at the end points is
\begin{equation}
   F(0)=V_p-\frac{4}{1+\lambda}>0,\quad  \lim_{r_0 \to 1} F(r_0)= -\infty.
\end{equation}
Then by the intermediate value theorem, it will have at least one zero located at $0<r_0<1$. The derivative of $F(r_0)$ is
\begin{equation}
F'(r_0)=-\left(\frac{4r_0\left(5+\lambda+(1+\lambda)r_0^2\right)}{(1+\lambda)(1-r_0^2)^3}\right)<0, \quad 0<r_0<1.
\label{eq: rdev}
\end{equation}
Then $F(r_0)$ is decreasing monotonic and can only have one zero in the range $0<r_0<1$.  $\square$
\subsection{Condition for a unique zero of the angular equation in \refeq{eq: polarred}} \label{sec: proofH}
Take the function 
\begin{equation}
  H(r_0)=\left(\frac{4-2(-1+\lambda)r_0^2+2(1+\lambda)r_0^4}{r_0(1+\lambda)(1-r_0^2)^3}-\frac{1}{r_0}V_p\right).
\end{equation}
 It will have the same zeros as 
 \begin{equation} G^\ast(r_0)=4+2(1-\lambda)r_0^2+2(1+\lambda)r_0^4-(1+\lambda)(1-r_0^2)^3V_p.
 \end{equation}
  $G^\ast(r_0)$ is reduced to a cubic equation with $r_0^2\equiv x$ yielding
 \begin{equation} G(x)=4+2(1-\lambda)x+2(1+\lambda)x^2-(1+\lambda)(1-x)^3V_p.
 \end{equation}
 \\
 
 \noindent \textbf{Proposition}: Assume $\lambda>0$ and the regime $V_p>\frac{4}{1+\lambda}$. Then $G(x)$ has an unique positive zero and it is located in the range $0<x<1$.
 \newline
\textbf{Proof}
The evaluation of the function at the end points is
\begin{equation}
   G(0)=4-(\lambda+1)V_p,\qquad G(1)=8>0. 
\end{equation}
If $V_p>\tfrac{4}{1+\lambda}$, then $G(0)<0$ while $G(1)>0$.
Then by the intermediate value theorem it will have at least one zero located at $0<r_0<1$.

The derivative of  $G(x)$ is
\begin{equation}
    G'(x)
=2(1-\lambda)+3(1+\lambda)V_p+(1+\lambda)(4-6V_p)x+3(1+\lambda)V_px^2.
\end{equation}
$G'(x)$ is a convex 2nd order polynomial as $3(1+\lambda)V_p>0$ with a minimum located at $x_{min}=1-\frac{2}{3V_p}$ with $G'(x_{min})=2(3+\lambda)-\frac{4(1+\lambda)}{3V_p}$.

If $V_p \geq 2/3$ then  $G'(x)>0,$  for all $x$. Then as $G(0)<0$ it can only have one positive zero and must be in the range $0<x<1$.
If  $V_p < 2/3$ then $x_{min}<0$. As $x_{min}<0$ then $G'(x)$ has at most one zero in the range $x>0$.
Consider the following two cases. First $G'(x)$ has a zero in the range $x>0$ then $G'(0)<0$. Then as  $G(0)<0$ and $G(1)>0$, $G'(x)$ must be positive somewhere in the range $0<x<1$,  thus the root for $G'(x)$ is between $0<x<1$. Since there can only be one sign change in $G'(x)$, $G(x)$ has exactly one positive zero and must be in the range $0<x<1$.
Second, if $G'(x)$ has no zeros or $G'(x)$ has two zeros for $x<0$ then $G'(x)>0$ thus $G(x)$ is a strictly increasing function and has exactly one positive zero and must be in the range $0<x<1$. $\square$
\\

Thus $G^\ast(r_0)$ has exactly one zero in the range $0<r_0<1$. Also then $H(r_0)$ has exactly one zero in the range $0<r_0<1$. Moreover $\lim_{r_0 \to 1^-} H(r_0)= \infty$. As the only zero of $H(r_p)=0$ in the range $0<r_0<1$ is $r_p$ then, 
\begin{equation}
    \begin{split}
     &H(r_0)<0,\quad r_0<r_p,\\ 
     &H(r_0)>0,\quad r_0>r_p.   
    \end{split}
    \label{eq: Hsign}
\end{equation}

\subsection{Linear stability analysis of the system in \refeq{eq: polarred}}
\label{sec: jacobian}
Consider the system
\begin{equation}
\begin{split}
&\frac{d r_0}{dt}= F(r_0)\cos(\theta_0),\quad F(r_0)\equiv\left(V_p-2\frac{2+(1+\lambda)r_0^2}{(1+\lambda)(r_0^2-1)^2}\right),\\
&\frac{d\theta_0}{dt}= H(r_0)\sin(\theta_0), \quad  H(r_0)\equiv \left(\frac{4-2(-1+\lambda)r_0^2+2(1+\lambda)r_0^4}{r_0(1+\lambda)(1-r_0^2)^3}-\frac{1}{r_0}V_p\right).
\end{split}
\end{equation}
It has fixed points  for $0<r_0<1$ at $(r_c,\theta_0)$=$(r_c,0)$,$(r_c,\pi)$,$(r_p,\pi/2)$ and $(r_p,3\pi/2)$. $r_c$ is the unique zero of $F(r_c)=0$ in the range $0<r_0<1$, shown in \refsec{sec: proofF}. $r_p$ is the unique zero of $H(r_p)=0$ in the range $0<r_0<1$, shown in \refsec{sec: proofH}.
The Jacobian of the system is 
\begin{equation}
    J(r_0,\theta_0)=\begin{bmatrix}
        F'(r_0)\cos(\theta_0)&&   -F(r_0)\sin(\theta_0)\\
        H'(r_0)\sin(\theta_0)&&   H(r_0)\cos(\theta_0)\\
    \end{bmatrix}.
\end{equation}
For the fixed points  at $(r_0,\theta_0)$=$(r_c,0)$ and $(r_c,\pi)$
\begin{equation}
    J(r_c,\theta_c)=\begin{bmatrix}
        \pm F'(r_c)&&   0\\
        0&&   \pm H(r_c)\\
    \end{bmatrix}.
\end{equation}
With the plus corresponding to the fixed point at $\theta_0=0$ and the minus to $\theta_0=\pi$.  
As the matrix is diagonal, the eigenvalues are going to be the diagonal entries.  
The eigenvalues are also real, since the matrix entries are real.  
The stability analysis is easier using the trace and determinant given by 
\begin{equation}
    \begin{split}
        &Tr(J(r_c,\theta_c))=\mp \frac{8(2+\lambda)r_c}{(1+\lambda)(1-r_c^2)^3},\\
        &D(J(r_c,\theta_c))= F'(r_c)H(r_c).
    \end{split}
\end{equation}
For the fixed point at $\theta_c=0$ the trace is negative, and for the fixed point at $\theta_c=\pi$ the trace is positive. $F'(r_c)<0$ as seen in \refeq{eq: rdev}.  For $H(r_c)$, if $r_c<r_p$ then $H(r_c)<0$, and if $r_p<r_c$ then $H(r_c)>0$, as shown in \refeq{eq: Hsign}.

For the fixed point at $(r_c,0)$, the trace is  $Tr(J(r_c,0))<0$.  
The case $r_c<r_p$ yields $D(J(r_c,0))>0$ thus the fixed point  is  a stable node. The case $r_c>r_p$  yields $D(J(r_c,0))<0$, then the fixed point is a saddle.  
For the fixed point at $(r_c,\pi)$, the trace is $Tr(J(r_c,\pi))>0$. The case $r_c<r_p$ yields $D(J(r_c,\pi))>0$ then the fixed point is an unstable node. For the case $r_c>r_p$, the determinant is $D(J(r_c,\pi))<0$, thus it is a saddle. 

For the fixed points  at $(r_0,\theta_0)=(r_p,\pi/2)$ and $(r_p,3\pi/2)$ the Jacobian is
\begin{equation}
    J(r_0,\theta_0)=\begin{bmatrix}
       0&&   \mp F(r_p)\\
       \pm H'(r_p)&&  0\\
    \end{bmatrix}.
\end{equation}
With the top sign corresponding to the fixed point at $\theta_0=\pi/2$ and the bottom sign to $\theta_0=3\pi/2$.  
$Tr(J)=0$, then fixed points  will be either saddles or centers depending on the sign of the determinant, $D(J)=H'(r_p)F(r_p)$.  
$H'(r_p)>0$ as seen in \refeq{eq: Hsign}, and $F(r)$ is a decreasing monotonic function as seen in \refeq{eq: rdev}, with $F(r_c)=0$.  
If $r_c<r_p$, then $F(r_p)<0$ and $D(J)<0$, and both fixed points will be saddles.  
If $r_p<r_c$, then $F(r_p)>0$ and $D(J)>0$, thus both fixed points will be centers according to linear stability analysis.  
To confirm that these are true centers, a conserved quantity leading to closed orbits must be identified.  
Refer to \refeq{eq: consQ} for the conserved quantity.


\end{document}